\documentclass[prd,twocolumn,superscriptaddress,showpacs,nofootinbib,preprintnumbers]{revtex4}

\usepackage{amsmath}
\usepackage{amsfonts}
\usepackage{graphicx}
\usepackage{dcolumn}
\usepackage{hyperref}

\def\be{\begin{equation}}
\def\ee{\end{equation}}
\def\ba{\begin{eqnarray}}
\def\ea{\end{eqnarray}}
\def\bs{\begin{subequations}}
\def\es{\end{subequations}}
\renewcommand{\S}{{\text{\tiny $\phi$}}}
\newcommand{\T}{{\text{\tiny $T$}}}

\newcommand{\sV}{{\text{\tiny $\phi V$}}}

\newcommand{\tV}{{\text{\tiny $TV$}}}

\newcommand{\eff}{{\text{\tiny eff}}}
\newcommand{\vp}{\varphi}

\begin{document}

\vspace{1cm}

\title{Observational constraints on patch inflation in noncommutative spacetime} 

\author{Gianluca Calcagni}
\email{calcagni@fis.unipr.it}
\affiliation{Dipartimento di Fisica, Universit\`{a} di Parma, Parco 
Area delle Scienze 7/A, I-43100 Parma, Italy}
\affiliation{INFN -- Gruppo collegato di Parma, Parco Area delle 
Scienze 7/A, I-43100 Parma, Italy} 
\author{Shinji Tsujikawa} 
\email{shinji@nat.gunma-ct.ac.jp}
\affiliation{Department of Physics, Gunma National College of 
Technology, Gunma 371-8530, Japan}
\date{July 27, 2004}

\begin{abstract}
We study constraints on a number of patch inflationary models in noncommutative 
spacetime using a compilation of recent high-precision observational data. 
In particular, the four-dimensional General Relativistic (GR) case, the Randall-Sundrum (RS) and 
Gauss-Bonnet (GB) braneworld scenarios are investigated by extending previous commutative analyses 
to the infrared limit of a maximally symmetric realization of the stringy uncertainty principle. 
The effect of spacetime noncommutativity modifies the standard consistency relation between the 
tensor spectral index and the tensor-to-scalar ratio. We perform likelihood analyses in terms of inflationary 
observables using new consistency relations and confront them with large-field inflationary models with potential 
$V \propto \vp^p$ in two classes of noncommutative scenarios.
We find a number of interesting results: (i) the quartic potential ($p=4$) is rescued from marginal 
rejection in the class 2 GR case, and (ii) steep inflation driven by an exponential potential ($p \to \infty$) 
is allowed in the class 1 RS case.  
Spacetime noncommutativity can lead to blue-tilted scalar and tensor spectra even for monomial potentials, 
thus opening up a possibility to explain the loss of power observed in the cosmic microwave background anisotropies.
We also explore patch inflation with a Dirac-Born-Infeld tachyon field and explicitly show 
that the associated likelihood analysis is equivalent to the one in the ordinary scalar field case 
by using horizon-flow parameters. 
It turns out that tachyon inflation is compatible with 
observations in all patch cosmologies even for large $p$.
\end{abstract}
\preprint{astro-ph/0407543}

\maketitle


\section{Introduction}

Early-Universe observations have come to the golden age. 
The first-year results of the Wilkinson Microwave Anisotropy Probe 
(WMAP) \cite{ben03,spe03,pei03} 
provided high-precision cosmological dataset from 
which early-Universe models can be tested.
The observations strongly support the inflationary paradigm
based on General Relativity (GR) as a backbone of high-energy physics.
In particular, nearly scale-invariant and adiabatic density perturbations generated 
in single-field inflation exhibit an excellent agreement with the
observed cosmic microwave background (CMB) anisotropies \cite{pei03,bri03,BLM,KKMR,LL,teg04}.
Together with the upcoming high-precision data by the Planck 
satellite \cite{planc}, it will be possible to discriminate between a host of 
inflationary models from observations.
 
From the theoretical side, there has been a lot of interest in the 
construction of high-energy models of the early Universe.
This interest is motivated by string cosmology, an attempt 
to merge string theory and cosmology as a 
generalization of the standard inflationary paradigm based on GR.
A well-known example is the Randall-Sundrum (RS) II braneworld 
scenario \cite{RS}, in which our four-dimensional brane is embedded 
in a five-dimensional bulk spacetime (see, e.g.,\,\cite{mar03,BVD,cal3}
for a list of references on this subject).
The spectra of perturbations generated in braneworld inflation are
modified due to the presence of the 5th dimension \cite{MWBH,LMW}, 
but the consistency 
equation relating the tensor-to-scalar ratio to the spectral index 
of tensor perturbations is unchanged \cite{HuL1,HuL2}.
Nevertheless, observational constraints in terms of underlying potentials
are different compared to the GR case \cite{LS,TL}.
It was shown that the quartic chaotic potential ($V \propto \phi^4$) 
is under a strong observational pressure and that steep inflation driven by 
an exponential potential is ruled out.
This situation changes if we account for the Gauss-Bonnet (GB) curvature 
invariant in five dimensions, arising from leading-order quantum 
corrections of the low-energy heterotic gravitational action
(see Ref.~\cite{LN} for one of the first works about GB braneworld inflation).
One effect of the GB term is to break the degeneracy of the standard consistency 
relation \cite{DLMS}. Although this does not lead to a significant change 
for the likelihood results of the inflationary observables, the quartic potential 
is rescued from marginal rejection for a wide range of energy 
scales \cite{TSM}. Even steep inflation exhibits marginal 
compatibility for a sufficient number of $e$-folds ($N \gtrsim 55$).

Another interesting string-inspired scenario is noncommutative 
inflation \cite{ABM,BH} in which the presence of a stringy 
spacetime uncertainty relation leads to the modification of 
perturbation spectra at large scales.
The implications of maximally symmetric noncommutativity 
have been extensively studied, e.g., in 
Refs.~\cite{HL1,FKM,FKM2,TMB,HL2,HL3,LiLi,KLM,KLLM,Cai,LMMP,AN,cal4,cal5,myu04}. 
Since the uncertainty relation is saturated when a perturbation with 
a particular wavelength is generated, the standard evolution of 
commutative fluctuations is altered and large-scale modes are suppressed.
In Ref.~\cite{TMB} the CMB power spectrum was divided into 
three main regions, ultraviolet (UV), intermediate, and infrared (IR).
It was also assumed that the perturbation spectra at low multipoles
are generated between the UV and intermediate regimes to 
explain the loss of power observed in CMB anisotropies.
In this case, however, the suppression of power is not significant 
since the intermediate spectrum is nearly scale invariant.
In this work we will adopt another perspective, that is, to 
consider the far IR regime as a dominant contribution to 
the large-scale spectrum.

We implement both braneworld and noncommutative frameworks 
as well as the standard GR commutative/noncommutative paradigm.
We place observational constraints on a number of models generated by 
different braneworld and noncommutative prescriptions. 
Using the patch 
framework \cite{cal3}\footnote{Observational constraints on 
commutative patch inflation have also been considered in \cite{KM}.} 
coupled to the slow-roll (SR) 
formalism, we will derive consistency equations between the
inflationary observables: the scalar 
($n_s$) and tensor ($n_t$) indices, their runnings ($\alpha_s$ and 
$\alpha_t$), and the tensor-to-scalar ratio $R$. 
In particular, the consistency relation 
between $n_t$ and $R$ is modified by the effect of 
spacetime noncommutativity, which breaks the degeneracy of 
the consistency relation in the commutative GR/RS cases.
We have a variety of consistency relations for each patch cosmology (see Table \ref{table2} in Sec. \ref{setup}).
Our numerical analysis based on recent 
observational data shows that the general shape of likelihood contours 
in the $n_s$-$R$ plane is deformed independently by braneworld and 
noncommutative effects.
The major modification to the GR/RS 
commutative cosmology appears for the upper bound of $R$, roughly 
setting it in a $2\sigma$ interval with $0.5 \lesssim R_\text{max} \lesssim 
0.7$. The scalar index always ranges $0.9<n_s<1.1$ at the 
$2\sigma$ level. 

We also place constraints on monomial potentials $V=V_0 \vp^p$ 
(including the exponential potential by taking the limit $p \to \infty$) 
in GR/RS/GB cases in noncommutative spacetime.
An intriguing feature is that the effect of the new physics
allows a blue-tilted spectrum, 
thus giving rise to a variety of theoretical points in the 
$n_s$-$R$ plane. For example, steep inflation driven by an exponential 
potential is excluded in the commutative RS scenario, but 
the same potential can be compatible with observations in the 
IR noncommutative limit. 
We shall investigate the observational compatibility of each 
scenario in detail.  
The suppression of the power spectrum 
at low multipoles will also be discussed by using the IR blue-tilted 
spectrum.

In addition to the ordinary scalar inflation, we will consider the 
Dirac-Born-Infeld tachyon field. 
Soon after the first proposal 
by Gibbons \cite{gib02},
it became clear that the cosmology based upon a rolling tachyon 
suffers from a number of problems, including 
a difficulty of reheating and 
a large amplitude of density perturbations, that can be 
traced back to some fine-tuning requirements on the parameters of the 
model \cite{CGJP,FKS,KL,SCQ,BBS,BSS,PS,rae04} (see also \cite{PCZZ}).
Lately it was shown in Ref.~\cite{GST} that the problem of large
density perturbations is solved by considering a small warp factor $\beta$
in a warped metric. In addition, the problem of reheating
is overcome by accounting for a negative cosmological constant 
which may appear by the stabilization of modulus fields \cite{KKLT}.

By the reasons presented above, it is premature to exclude tachyon inflation.
In this work we will show that the likelihood analysis for the inflationary 
observables in the tachyon case is identical to that for the ordinary scalar 
field by expressing the observables in 
terms of horizon-flow parameters.
We also place observational constraints on the large-field
potentials in noncommutative GR/RS/GB cases.
Remarkably, even steep inflation is within the $2\sigma$ contour 
bound in all cases because of a small tensor-to-scalar ratio.

The paper is organized as follows.
In Sec.\,\ref{setup} we provide our general 
(non)commutative patch setup. 
In Sec.\,\ref{like1} we perform 
likelihood analysis using the latest observational data set.
After computing the theoretical values of $n_s$ and $R$
for the potential $V=V_0\vp^p$ in Sec.\,\ref{models}, these are 
confronted with the likelihood contour bounds in 
Sec.\,\ref{like2}. We give conclusions and discussion 
in Sec. \ref{disc}.


\section{The models} \label{setup}

Consider a four-dimensional Friedmann-Robertson-Walker (FRW)
cosmological background in which 
the effective Friedmann equation is given by 
\be 
\label{Hubble}
H^2=\beta_q^2 \rho^q\,,
\ee
where $H$ is the Hubble rate and $\beta_q$ and $q$ are
 constants. General relativity, Randall-Sundrum and Gauss-Bonnet braneworld 
cases correspond to $q=1$, $q=2$ and $q=2/3$, respectively.
We shall investigate a situation in which a homogeneous scalar field, $\vp$, is confined on the 3-brane. 
The field $\vp$ plays the role of the inflaton which is responsible for the generation of primordial perturbations.

In this paper we are interested in two candidates for the inflaton $\vp$.
The first one is a standard 
minimally coupled scalar field, $\phi$,
whose energy density is 
\be \label{rho}
\rho=\frac12 \dot{\phi}^2+V(\phi)\,,
\ee
where $V(\phi)$ is the potential of $\phi$. 
The inflaton satisfies 
the following equation of motion
\be \label{phieq}
\ddot{\phi}+3H\dot{\phi}+V'=0\,,
\ee
where a prime denotes the $\phi$ derivative. 
The second one is a
tachyon-type scalar field, $T$,  
with a Born-Infeld action. 
In this case the energy density is given by 
\be
\rho = \frac{V(T)}{\sqrt{1-\dot{T}^2}}\,,\label{Trho}
\ee
together with the equation of motion
\be \label{Teom}
\frac{\ddot{T}}{1-\dot{T}^2}+3H\dot{T}+U'=0\,,
\ee
where $U \equiv \ln V$ is differentiated with respect to $T$. 

The SR parameters 
associated with the inflaton $\vp$ are \cite{cal3}
\bs\ba
\epsilon_\vp &\equiv& 
-\frac{\dot{H}}{H^2}\,, 
\label{SR}\\
\eta_\vp  &\equiv&  -\frac{d \ln \dot{\vp}}
{d \ln a}=-\frac{\ddot{\vp}}{H\dot{\vp}} \label{Heta}\,,\\
\xi_\vp^2    &\equiv&   \frac{1}{H^2} 
\left(\frac{\ddot{\vp}}{\dot{\vp}}\right)^\cdot \,,\label{Hxi}
\ea\es
where $\vp=\phi, T$.
For example, in the case of the normal scalar field $\phi$, 
we have
\bs\ba 
\label{epphi}
\epsilon_\phi =
\frac{3q\beta_q^{2-\theta}}{2} 
\frac{\dot{\phi}^2}{H^{2-\theta}}\,,
\ea\es
where 
\be 
\label{theta}
\theta=2\left(1-\frac{1}{q}\right)\,,
\ee
while for the tachyon
\bs\ba 
\epsilon_T=
\frac{3q}{2}\dot{T}^2\,.
\ea\es
Then $\epsilon_\vp$ can be written as
\bs\ba 
\epsilon_\vp=
\frac{3q\beta_q^{2-\tilde{\theta}}}{2} 
\frac{\dot{\vp}^2}{H^{2-\tilde{\theta}}}\,,
\ea\es
where $\tilde{\theta}=\theta$ for $\vp=\phi$ and 
$\tilde{\theta}=2$ for $\vp=T$. 

As we shall see below, it is convenient to 
introduce the horizon-flow (HF) parameters \cite{STG,kin02}, 
defined by 
\be
\epsilon_0=\frac{H_{\rm inf}}{H}\,,\qquad
\epsilon_{i+1}=\frac{ d \ln |\epsilon_i|}{dN}\,,\qquad
i \ge 0\,,
\label{hflow}
\ee
where $H_{\rm inf}$ is the Hubble rate at some chosen time and $N 
\equiv \ln (a/a_i)$ is the number of $e$-folds; here $t_i$ 
is the time when inflation begins.\footnote{\label{foot1} Note that our 
definition, which counts $N$ forward in time, is in accordance 
with \cite{STG}, where $N(t_i)=0$ and goes up to $N(t)>0$. 
This is in contrast with the ``backward'' definition of \cite{kin02}, 
where $N=\ln (a_f/a)$ is the number of remaining $e$-folds at the time 
$t$ before the end of inflation at $t_f$. 
In Sec.\,\ref{models} we will adopt the backward notation; 
this is consistent since the change of definition involves completely independent 
analyses.} 
The evolution equation for the HF parameters is given by  
$\dot{\epsilon}_i = H\epsilon_{i}\epsilon_{i+1}$. 
The HF parameters are related to the first SR 
parameters, as
\bs\ba
\epsilon_1 &=& \epsilon_\vp\,,\\
\epsilon_2 &=& \big(2-\tilde{\theta}\big)
\epsilon_\vp-2\eta_\vp\,,\\
\epsilon_2\epsilon_3 &=& 
\big(2-\tilde{\theta}\big)^2
\epsilon_\vp^2-2\big(3-\tilde{\theta}\big)
\epsilon_\vp\eta_\vp+2\xi_\vp^2\,.
\ea\es

Noncommutative inflation arises when we impose a realization of the 
*-algebra on the brane coordinates; an algebra preserving the maximal 
symmetry is $[\tau,x]=il_s^2$, where $\tau \equiv \int a\,dt$, $x$ is 
a comoving spatial coordinate on the brane and $l_s \equiv M_s^{-1}$ is 
the 
fundamental string scale. Let us introduce the noncommutative 
parameter $\delta \equiv (M_s/H)^2$; the noncommutative algebra 
induces a cutoff $k_0(\delta)$ roughly dividing the space of comoving 
wave numbers into two regions, one encoding the UV, small-scale 
perturbations generated inside the Hubble horizon ($H \ll M_s$) and 
the other describing the IR, large-scale perturbations created 
outside the horizon ($H \gg M_s$). By definition, they correspond to 
the quasicommutative and strongly noncommutative regimes, 
respectively. It turns out that one can write the spectra of scalar 
and tensor perturbations in the form
\be 
\label{Anoncom}
A(\delta,\,H,\,\vp) = A^{(c)} (H,\,\vp)\,\Sigma (\delta)\,,
\ee
where $A^{(c)}=A(\Sigma\!\!=\!\!1)$ is the amplitude in the 
commutative limit and $\Sigma(\delta)$ is a function encoding 
leading-SR-order noncommutative effects. 

Equation (\ref{Anoncom}) is 
evaluated at the horizon crossing in the UV limit and at the time when 
the perturbation with comoving wavenumber 
$k$ is generated in the IR limit. 
To lowest order in the HF (or SR) parameters,
\be \label{dotsig}
\frac{d \ln \Sigma^2}{d \ln k}
= \sigma \epsilon_1\,,
\ee
where $\sigma = \sigma(\delta)$ is a function of $\delta$ such that 
$\dot{\sigma}=O(\epsilon_1)$. The standard commutative spacetime corresponds to $\sigma=0$.

The amplitudes of scalar and tensor perturbations are given, 
respectively, by 
\ba
\label{ampli}
A_s^2(\vp)  &=& 
\frac{3q\beta_q^{2-\theta}}{25\pi^2}
\frac{H^{2+\theta}}{2\epsilon_1} \Sigma^2\,, \\
A_t^2(\vp)  &=& 
\frac{3q\beta_q^{2-\theta}}{25\pi^2}
\frac{H^{2+\theta}}{2\zeta_q} \Sigma^2\,,
\ea
where the coefficient $\zeta_q$ in $A_t^2$ depends on the model one considers.
The standard GR case corresponds to $\zeta_1=1$, whereas
one has $\zeta_2=2/3$ for the RS braneworld \cite{LMW}.
In the GB case $\zeta_{2/3}=1$,
as recently shown in Ref.~\cite{DLMS}.

The spectral indices of scalar and tensor perturbations are given by
\ba
\label{ns}
n_s-1&\equiv& \frac{d \ln A_s^2}{d \ln k} = 
-(2+\theta-\sigma)\epsilon_1-\epsilon_2\,, \\
n_t &\equiv& \frac{d \ln A_t^2}{d \ln k}=
-(2+\theta-\sigma)\epsilon_1\,,
\label{nt}
\ea
where we used $d\ln\, k \approx Hdt$; the effect of noncommutativity is 
encoded in the $\sigma$ term. 
The ratio of tensor-to-scalar perturbations is
\be \label{ratio}
R \equiv 16 \frac{A_t^2}{A_s^2}
=\frac{16\epsilon_1}{\zeta_q}\,.
\ee
Since the same factor $\Sigma$ multiplies both the 
tensor and scalar amplitudes, their ratio is unchanged with respect to
the commutative case. 
We have $R=16\epsilon_1$ for the GR and GB cases and 
$R=24\epsilon_1$ for the RS case.
The runnings of the spectral indices, 
$\alpha_{s,t} \equiv d n_{s,t}/d\ln k$, 
are given by
\ba
\alpha_s &=& 
-(2+\theta-\sigma)\epsilon_1\epsilon_2-\sigma\bar{\sigma}
\epsilon_1^2-\epsilon_2\epsilon_3\,,\\
\alpha_t &=& 
-(2+\theta-\sigma)\epsilon_1\epsilon_2-
\sigma\bar{\sigma}\epsilon_1^2\,,
\ea
where $\bar{\sigma}$ is defined as $\bar{\sigma} \equiv 
-\dot{\sigma}/(H\sigma\epsilon)$.
If $\sigma$ is a constant, $\bar{\sigma}$ vanishes.
         
All the previous expressions can be combined into the (nonclosed) set 
of consistency equations:
\ba
n_t &=& -(2+\theta-\sigma) \frac{R\zeta_q}{16}\,,\label{ntconeq}\\
\alpha_s &=& \frac{R\zeta_q}{16} 
\left\{(2+\theta-\sigma)(n_s-1)\vphantom{\frac{1}{1}}\right.\nonumber\\
&&\quad\quad+\left.\left[(2+\theta-\sigma)^2-\sigma \bar{\sigma} 
\right]\frac{R\zeta_q}{16} 
\right\}-\epsilon_2\epsilon_3\,,\label{alps}\\
\alpha_t &=& \frac{R\zeta_q}{16} 
\left\{(2+\theta-\sigma)(n_s-1)\vphantom{\frac{1}{1}}\right.\nonumber\\
&&\quad\quad+\left.
\left[(2+\theta-\sigma)^2-\sigma \bar{\sigma} 
\right]\frac{R\zeta_q}{16} \right\}\,.
\ea
Notably these relations do not depend on which inflaton field 
one is assuming on the brane.
This means that the likelihood analysis for the field $\phi$
in terms of the variables 
$n_s$, $R$ and $\epsilon_3$ with given values of $q$, $\sigma$ and 
$\bar{\sigma}$ is identical to the one for the tachyon $T$.
Using the SR parameters given in (\ref{SR})--(\ref{Hxi}), 
the running of the scalar perturbations would split into 
two different equations \cite{cal4}:
\ba
\alpha_s(\phi) &=& \frac{R\zeta_q}{16} 
\left\{\vphantom{\frac{1}{1}}(5-\sigma)(n_s-1)+[4(3+\theta)\right.
\nonumber\\
&&\quad\qquad-\left.\sigma 
(7+\theta+\bar{\sigma}-\sigma)]\frac{R\zeta_q}{16} 
\right\}-2\xi_\S^2,\label{split1}\\
\alpha_s(T) &=& \frac{R\zeta_q}{16} 
\left\{\vphantom{\frac{1}{1}}(3+\theta-\sigma)(n_s-1)+[(2+\theta)
(3+\theta)\right.\nonumber\\
&&\quad\qquad\left.-\sigma 
(5+2\theta+\bar{\sigma}-\sigma)]\frac{R\zeta_q}{16}\right\}-2\xi_\T^2;\label{split2}
\ea
then one would need to perform two separate likelihood analyses.
In this sense, the HF parameters are a more convenient choice 
for numerical purposes than the SR parameters.

The consistency relation (\ref{ntconeq}) yields $n_t=-(2+\theta)\zeta_q 
R/16$ in the commutative spacetime ($\sigma=0$).
Then we have $n_t=-R/8$ for the GR and RS cases, while the GB case, corresponding to $n_t=-R/16$, 
breaks this degeneracy \cite{DLMS}.

One can ``deform'' the consistency equations by taking the
spacetime noncommutativity into account ($\sigma \ne 0$).
Two classes of noncommutative models have 
been found in the infrared region \cite{BH}.
In the first one (class 1), the FRW 2-sphere is 
factored out in the measure of the effective 4D perturbation action 
$z_k$, which is given by the product of the commutative measure $z$ 
and a correction factor of the $1+1$ noncommutative model. In the 
class 2 choice, the scale factor in the measure is everywhere 
substituted by an effective scale $a_\eff$ whose time dependence is 
smeared out by the nonlocal physics; since $z \propto a$, then $z_k=z 
a_\eff/a$. Inequivalent prescriptions on the ordering of the 
*-product in the perturbation action further split these two classes, 
but in the IR limit they give almost the same predictions 
\cite{cal4}. 

Since $\sigma$ is a nontrivial function of the noncommutative 
parameter, the space of the cosmological observables is enlarged, 
thus modifying the determination of the constraints from 
observations. However, in the far IR limit this function approaches a 
constant value, $\sigma=6$ in class 1 models and $\sigma=2$ in class 
2 models. 
For positive values of $\sigma$ the perturbation spectra tend to be blue tilted in the IR 
region relative to the UV commutative case.\footnote{It is 
interesting to note that the sign of the correction to the scalar 
index ($+$) and its running ($-$) agrees with the results coming from 
a pure spacial realization of the noncommutative algebra 
\cite{LMMP,AN}.} 
For example, in the GR class 1 IR case ($\theta=0$ and $\sigma=6$), 
we have $n_s-1=4\epsilon_1-\epsilon_2$ and $n_t=4\epsilon_1>0$
by Eqs.~(\ref{ns}) and (\ref{nt}), which means that 
the spectrum of gravitational waves is always blue tilted.
Tables \ref{table1} and \ref{table2} summarize the scenarios we are 
going to explore.
For later convenience we dub the 
commutative spacetime ($\sigma=0$) as ``class 0.''
We find that the quantity $n_t/R$ is always positive
in the class 1 IR case. Thus, the introduction of noncommutativity breaks the standard GR commutative 
consistency relation $n_t=-R/8$ in a variety of ways. 
\begin{table}[ht]
\caption{\label{table1}The values of $q$, $\theta$ and $\zeta_q$
for each patch inflation.}
\begin{ruledtabular}
\begin{tabular}{c|dcc}
          &  $GR$  &  $RS$   &  $GB$     \\ \hline
$q$       &    1   &    2    &   2/3     \\
$\theta$  &    0   &    1    &   $-1$    \\ 
$\zeta_q$ &    1   &  2/3    &    1      \\
\end{tabular}\end{ruledtabular}
\end{table}

\begin{table}[ht]
\caption{\label{table2}The consistency equation (\ref{ntconeq}) in 
the commutative UV and noncommutative IR limit.}
\begin{ruledtabular}
\begin{tabular}{c|dddd}
(Non)commutative   & \sigma &  \multicolumn{3}{c}{$n_t/R$}\\
models             &        &  $GR$  &  $RS$  &  $GB$   \\ \hline
Class 0 UV         &    0   &  -1/8  &   -1/8 & -1/16   \\
Class 1 IR         &    6   &   1/4  &   1/8  & 5/16   \\ 
Class 2 IR         &    2   &    0   &  -1/24 & 1/16   \\
\end{tabular}\end{ruledtabular}
\end{table}


\section{Likelihood analysis for noncommutative inflation} 
\label{like1}

In Ref.~\cite{TMB}, a large-scale power spectrum 
was derived in the context of power-law inflation
for an IR mode generated around $\tau \approx k_0l_s^2$.
If this spectrum corresponds to the scale 
around $1<\ell\lesssim 10$, then the cosmologically relevant modes 
with $10\lesssim\ell\lesssim 1000$ also belong to the same IR spectrum.
This is because the characteristic scale of this spectrum is 
$k_{s3}=10^{-5}k_{s2}$, where $k_{s2}$ is an intermediate scale 
around which the IR description becomes invalid
[see Eq.~(12) of \cite{TMB}].
In Ref.~\cite{TMB} it was assumed that the intermediate spectrum 
dominates 
at largest scales ($1<\ell\lesssim 10$), following the approach 
of Ref.~\cite{HL1}.
In this work we shall investigate a situation in which 
the IR spectra $\Sigma^2 \sim 
\delta^3$ [class 1, which is Eq.~(23) of Ref.~\cite{TMB} in the de Sitter 
limit] and $\Sigma^2 \sim \delta$ (class 2) correctly describe the 
large-scale sector with $1<l \lesssim 10$. 
Since one generally has $k_{s2} \gg k_{s3}$,
it is natural to use the IR power spectrum over the cosmologically 
relevant scales with $1<\ell\lesssim 2000$.

We have run the Cosmological Monte Carlo CosmoMC code together 
with the CAMB program \cite{LCS,LB,camb}, applied to the latest 
observational data coming from the dataset of WMAP \cite{wmap}, 2dF 
\cite{per01} and SDSS \cite{spe03,teg04}. 
We implement the band-powers on small scales ($800 \lesssim l \lesssim 
2000$)
coming from CBI \cite{CBI}, VSA~\cite{VSA} and
ACBAR~\cite{ACBAR} experiments.

The set of inflationary observables is 
$\{A_s^2,R,n_s,n_t,\alpha_s,\alpha_t,\sigma\}$. 
The tensor index is absorbed via the consistency equation (\ref{ntconeq}) 
while $\alpha_t$ is ignored since its cosmological impact is too 
small to be detected in current observations.
The actual set of parameters is 
$\{A_s^2,\epsilon_1,\epsilon_2,\epsilon_3,\sigma\}$
or equivalently $\{A_s^2, n_s, R, \epsilon_3,\sigma\}$.
For several fixed values of $\sigma$ 
($\sigma=0, 2, 6$)
we have numerically found that 
$\epsilon_3$ is poorly constrained and is consistent to be set to zero. 
We also ran the numerical code when the SR parameters $\epsilon$, $\eta$
and $\xi$ are varied. 
Since the running $\alpha_s$ is constrained to be $|\alpha_s| 
\lesssim 0.03$ in order for the Taylor expansion of the power 
spectrum to be valid \cite {LL}, one cannot put large values of the prior 
on $\xi^2$. Making use of the fact that $\xi^2$ is of the same 
order as $\epsilon_1 \epsilon_2$ and the two HF parameters are
constrained to be $\epsilon_1 \lesssim 0.03$ 
and $|\epsilon_2| \lesssim 0.1$ \cite{LL}, we should put the prior 
around $\xi^2<0.003$. 
In this case our likelihood analysis shows that $\xi^2$
vanishes consistently.
We also numerically found that the likelihood values of 
inflationary and cosmological parameters are very  
similar to the case in which the HF parameters are used. 
In what follows we shall show the numerical results obtained 
by using HF parameters, since this is more convenient 
because of the degeneracy between the ordinary field $\phi$
and the tachyon field $T$.\footnote{Although the available data are compatible 
with both the assumptions $|\epsilon_3| \ll \epsilon_1$ and $|\xi| \ll \min (\epsilon,|\eta|)$ 
(this one adopted in \cite{cal3,cal4}), there may be some difference between 
the HF and SR towers from a purely theoretical point of view, when considering
the issue of degeneracy among cosmological models. In fact, within the
slow-roll formalism, the consistency equation for the scalar running splits
into Eqs. (\ref{split1}) and (\ref{split2}) and the HF degeneracy of ordinary
scalar and tachyon scenarios is removed.
Therefore the two choices one imposes to close the equation of the scalar 
running in terms of observables lead to different consequences 
which are outlined in \cite{cal5}.} 

\begin{figure}
\includegraphics[height=3.5in,width=3.5in]{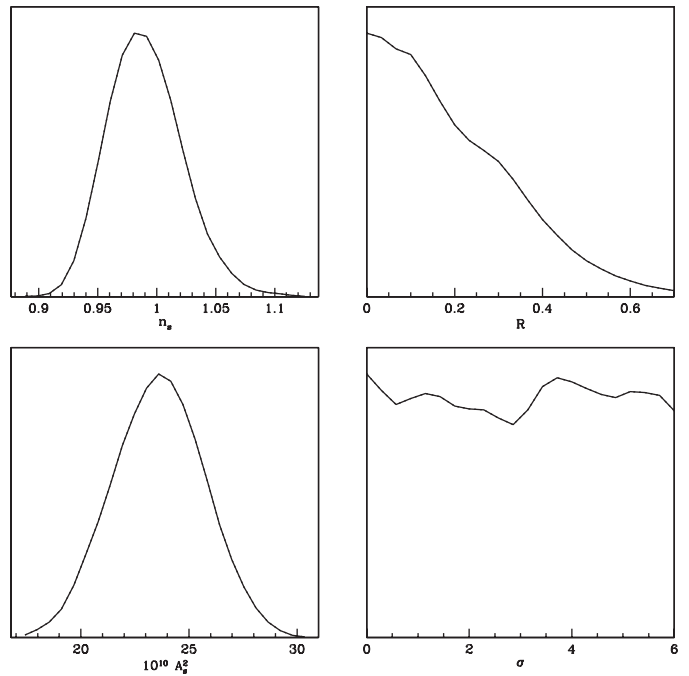}
\caption{\label{fig1}
Marginalized probability distributions 
of inflationary parameters ($n_s$, $R$, $A_s^2$, $\sigma$)
for the GR case with prior $0 \le \sigma \le 6$ 
and $\epsilon_3=0$. The likelihood analysis does not 
choose a preferred value of the
noncommutative parameter $\sigma$.
}
\end{figure}

We ran the numerical code for the GR case by varying 
$A_s^2$, $n_s$, $R$ and $\sigma$ with $\epsilon_3=0$. 
We chose the parameter range $0 \le \sigma \le 6$
including the commutative limit ($\sigma=0$)
and two classes of noncommutative IR limits ($\sigma=6, 2$).
In the GR case the consistency relation (\ref{ntconeq})
reads $n_t=-(2-\sigma)R/16$, which means that the
ratio $n_t/R$ ranges $-1/8 \le n_t/R \le 1/4$.
As found in  Fig.~\ref{fig1},
$\sigma$ does not select a preferred 
value, since the $R=0$ case is not ruled out anyway. 
Therefore noncommutative inflation is allowed
observationally as well as the case of commutative 
spacetime.

Since the UV commutative case has already been investigated in 
literature, we will concentrate ourselves to the IR noncommutative 
region. This choice is also dictated by a technical reason. 
In the commutative limit $\sigma$ depends upon both the 
Hubble parameter, evaluated at the horizon crossing, 
and the string mass $M_s$. 
As we have seen, even if the tensor spectral index is fixed 
by Eq.~(\ref{ntconeq}), the introduction 
of the extra degree of freedom 
$\sigma$ results in a poor constraint on the parameter itself.
On the contrary, in the far IR region the function $\sigma$ approaches 
nonzero constant values as shown in Table \ref{table2}. 
This allows us to impose the consistency equation (\ref{ntconeq}) 
and concretely reduce the space of parameters, setting a 
meaningful scheme 
of analysis for the noncommutative models. Moreover, the amplitude of 
gravitational waves is strongly damped for angular scales with 
$\ell \gtrsim 10$ and the relations (\ref{ratio}) and (\ref{ntconeq}) 
only affect the large scales with $\ell \lesssim 10$, corresponding 
to the IR region. In this sense, using a constant $\sigma$ is a good 
approximation.

\begin{figure}
\includegraphics[height=3.5in,width=3.5in]{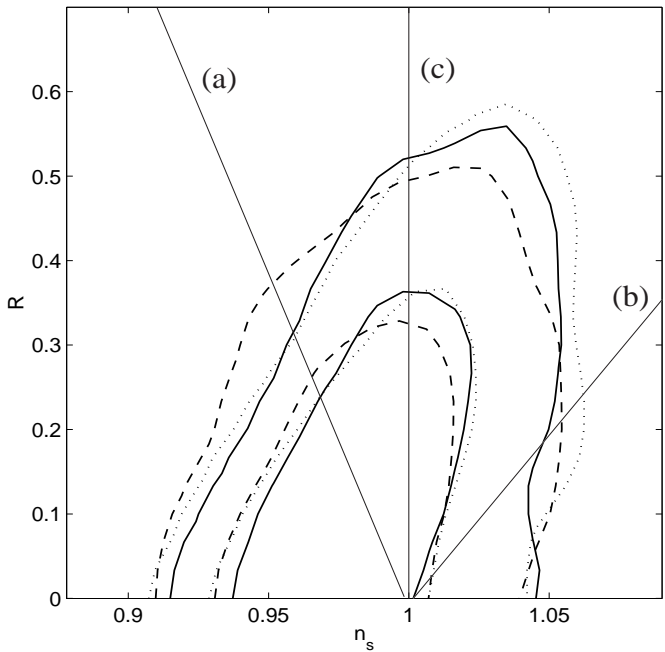}
\caption{\label{fig2}
The $1\sigma$ and
$2\sigma$ observational contour bounds for the GR case.
Each contour curve corresponds to (\textit{a}) GR0 ($\sigma=0$), 
solid line; (\textit{b}) GR1 ($\sigma=6$), 
dashed line; (\textit{c}) GR2 ($\sigma=2$), dotted line.
We also show the border of large-field and hybrid 
inflationary models for (\textit{a}) GR0, (\textit{b}) GR1 
and (\textit{c}) GR2 cases.  
The region on the left of each border corresponds to the parameter
space in large-field models.
Noncommutative spacetime allows the border
extending to the region $n_{s}>1$.
}
\end{figure}

\begin{figure}
\includegraphics[height=3.5in,width=3.5in]{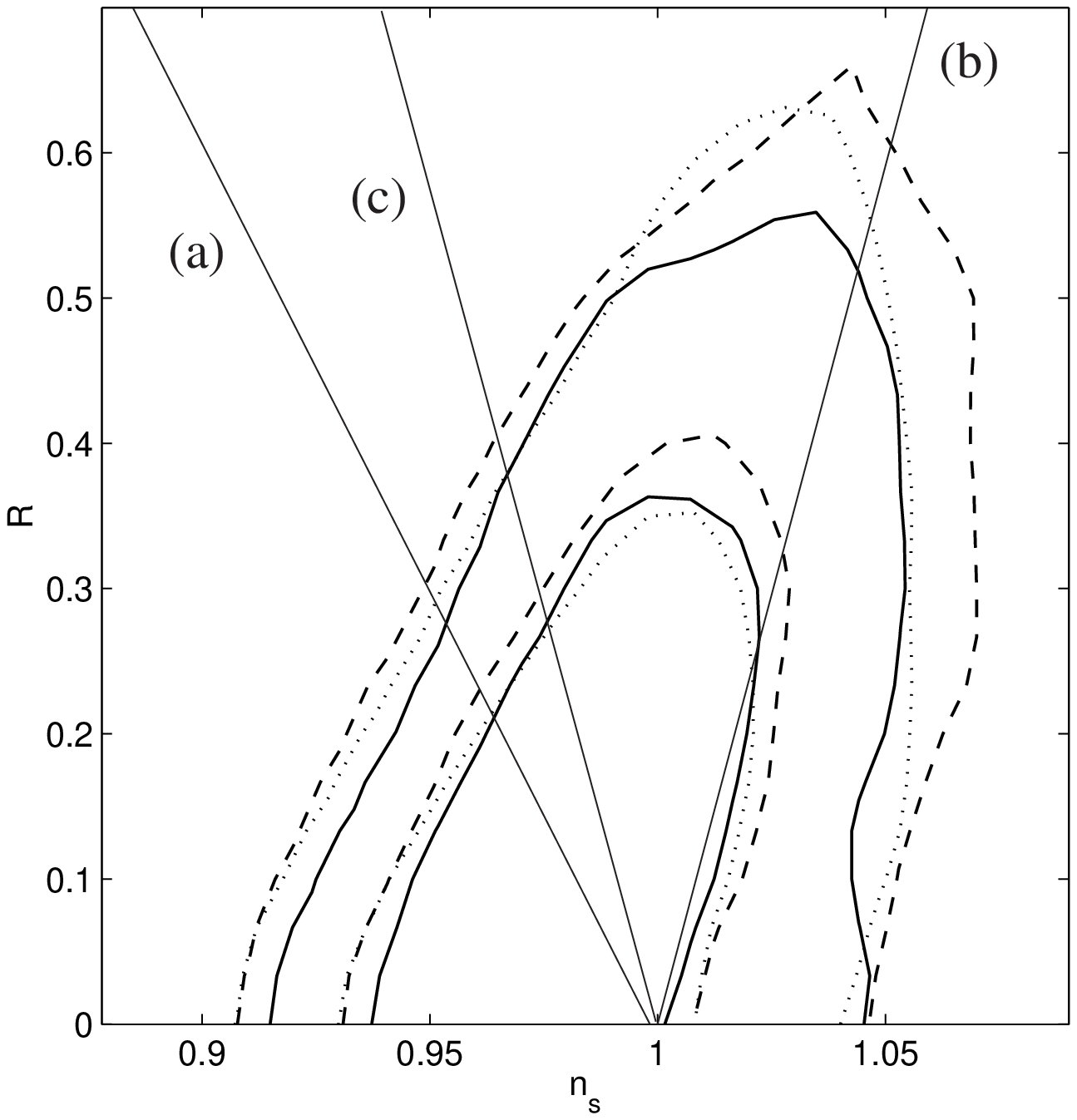}
\caption{\label{fig3}
The $1\sigma$ and $2\sigma$ observational contour bounds for the RS case.
The meaning of the curves and the borders are 
the same as in Fig.~\ref{fig2}.
}
\end{figure}

\begin{figure}
\includegraphics[height=3.5in,width=3.5in]{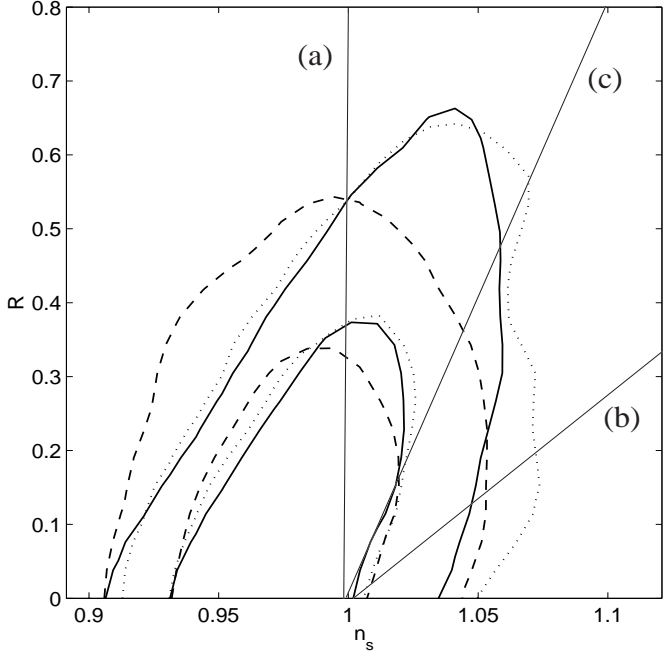}
\caption{\label{fig4}
The $1\sigma$ and
$2\sigma$ observational contour bounds for the GB case.
The meaning of the curves and the borders are 
the same in Fig.~\ref{fig2}.
}
\end{figure}

In Fig.~\ref{fig2} we plot the $1\sigma$ and
$2\sigma$ observational contour bounds for the GR case
with $\sigma=0$ (GR0), $\sigma=6$ (GR1) and 
$\sigma=2$ (GR2).
Figures \ref{fig3} and \ref{fig4} correspond to the 
likelihood contours for the RS and GB cases, 
respectively.\footnote{The likelihood contour for the Gauss-Bonnet 
case is slightly different from what was obtained in Ref.~\cite{TSM}. 
This is because in that paper the authors considered 
the exact GB scenario and assumed that the running of the 
spectral indices is zero, since the expressions of the 
exact RS and GB regimes are very complicated.
This resulted in a lower upper bound 
for $R$.}
These results hold for both the scalar field $\phi$ and 
the tachyon field $T$ because of the use of the 
HF parameters.

In the GR case, the class 2 ($\sigma=2$) is rather special since $R$ 
and $n_t$ vanish. 
The class 2 contour extends to higher 
values of $R$ relative to the commutative plot, while the class 1 
contour allows larger values of $|n_s-1|$ but with a smaller 
$R_{\rm max}$. Thus the noncommutativity of a model is not 
monotonically measured by $\sigma$ (with greater $\sigma$ 
corresponding to larger effects) and nonlocal features make their 
appearance in a nontrivial way.

We can do similar considerations for the RS case (where the maximal 
elongation is achieved for $\sigma=6$) and for the GB one (where 
the class 1 behaves in a totally different manner); see 
Figs.~\ref {fig3} and \ref {fig4}.
 Note that the 
degeneracy between GR and RS is removed for $\sigma>0$, both from a 
theoretical and observational point of view.


\section{Theoretical values of inflationary observables in 
large-field noncommutative models} \label{models}

While the HF equations of Sec.\,\ref{setup} are independent of the 
kind of inflaton when expressed via the horizon-flow parameters, 
the difference appears when we constrain inflationary potentials.
This can be clearly understood in the SR formalism, using the 
potential SR tower $\{\epsilon_\sV,\eta_\sV,\xi_\sV,\dots\}$ 
described in \cite{cal3}, together with the first-order relations 
$\epsilon_\S \approx \epsilon_\sV$, $\eta_\S \approx 
\eta_\sV-\epsilon_\sV$, $\xi_\S^2 \approx 
\xi_\sV^2-3\epsilon_\sV\eta_\sV+(3-\theta)\epsilon_\sV^2$. Even if 
the dynamical conditions are slightly more precise within the Hubble 
SR formalism, the potential approximation fits better for our 
analysis. The inflationary observables $A_s^2$, $n_s$ and 
$R$  read, to lowest SR order,
\bs\ba
A_s^2(\phi) &=& 
\frac{9\beta_q^6}{25\pi^2}\frac{V^{3q}}{V'^2}\Sigma^2\,,\\
n_s-1  &=& 2\eta_\S-(4-\sigma)\,\epsilon_\S\\
&=&\frac{1}{3\beta_q^2V^q}\left[2V''+(\sigma-6)
\frac{q}{2}\frac{V'^2}{V}\right]\,,\\
R &=& \frac{16q}{6\beta_q^2\zeta_q}\frac{V'^2}{V^{q+1}}\,.
\ea\es
For the tachyon, $\epsilon_\T \approx \epsilon_\tV$, $\eta_\T \approx 
\eta_\tV$, $\xi_\T^2 \approx \xi_\tV^2+3\epsilon_\tV\eta_\tV$, and 
the inflationary observables are
\bs\ba 
A_s^2(T) &=& 
\frac{9\beta_q^6}{25\pi^2}\frac{V^{3q+1}}{V'^2}\Sigma^2\,,\\
n_s-1    &=& 2\eta_\T-(2+\theta-\sigma)\,\epsilon_\T\\
         &=& 
\frac{1}{3\beta_q^2V^q}\left[2U''-(4+\theta-\sigma)
\frac{q}{2}U'^2\right]\,,\\
R &=& \frac{16q}{6\beta_q^2\zeta_q} \frac{V'^2}{V^{q+2}}\,.
\ea\es

In this and the following section we shall consider an important 
class of inflaton potentials, namely, the large-field models
\be \label{power}
V(\vp)=V_0 \vp^p\,,\qquad \vp=\phi,\,T\,,
\ee
in which the inflaton field starts with a large initial value and rolls
down toward the potential minimum at smaller $\vp$. 
The linear potential  with $p=1$ corresponds to 
the border of large-field and 
small-field models.
The exponential potential
\be \label{expo}
V=V_0\exp\left(-\vp/\vp_0\right)\,,
\ee
characterizes the border of large-field and hybrid models.
This case can be regarded as the $p \to \infty$ limit of 
the polynomial potential (\ref{power}).

Making use of slow-roll approximations 
$\big|\ddot{\phi}\big| \ll \big|3H\dot{\phi}\big|$ and $\dot{\phi}^2 \ll V(\phi)$, 
we obtain  
\be
N(\phi) \approx -3\beta_q^2
\int_\phi^{\phi_f}\frac{V^q}{V'}\,d\phi\,,
\label{efold1}
\ee
for the scalar field $\phi$, and
\be
N(T) \approx -3\beta_q^2
\int_T^{T_f}\frac{V^{q+1}}{V'}\,dT\,,
\label{efold2}
\ee
for the tachyon field $T$. Here we make use of the backward 
definition $N=\ln (a_f/a)$ of the number of $e$-foldings; see footnote \ref{foot1}.

The potentials (\ref{power}) and (\ref{expo}) cover a number of exact 
solutions either exactly or approximately \cite{cal3}. In fact, the 
commutative solutions of Ref.~\cite{cal3} are perfectly viable in the 
noncommutative case too, since the nonlocal physics does not affect 
the homogeneous background. The unique apparently subtle point is 
that in the IR region one explicitly uses the exponential solution to 
construct the perturbation amplitudes, contrary to the UV case in 
which it is implicitly assumed in the approximation of constant SR 
parameters. However, the subtended philosophy is 
quite the same, that is 
to find a general solution with constant nonzero SR parameters and 
then to perturb it with small time variations. Despite these 
simple considerations, the predictions of these homogeneous models 
definitely change when spacetime becomes noncommutative.


\subsection{The ordinary scalar field $\phi$}

For the scalar potential (\ref{power}) with the ordinary 
scalar field $\phi$, we have 
\ba
n_s-1 &=& -\frac{pV_0^{1-q}}{6\beta_q^2}\frac{p(6q-\sigma 
q-4)+4}{\phi^{2+(q-1)p}}\,,\label{npow}\\
R &=& \frac{16qp^2}{6\beta_q^2\zeta_qV_0^{q-1}}\phi^{(1-q)p-2}\,.
\ea
We can estimate the field value at the end of inflation 
by setting $\epsilon_\phi(t_f)=1$, which yields 
$\phi_f^{p(q-1)+2} \approx qp^2/
(6\beta_q^2V_0^{q-1})$.\footnote{One may adopt the criterion 
$\eta_\sV(t_f)=1$ to estimate the value $\phi_f$, but the difference 
is small as long as $p/N \ll 1$.}

Then the number of $e$-foldings (\ref{efold1}) is
\ba \label{efolds}
N =\frac{3\beta_q^2V_0^{q-1}}{p[p(q-1)+2]}\,\phi^{p(q-1)+2}-
\frac{qp}{2[p(q-1)+2]}\,,
\ea
which is valid for $p \ne 2/(1-q)$.
The scalar index and the tensor-to-scalar ratio are
\ba
n_s-1 &=& -\frac{p(6q-\sigma q-4)+4}{2N(pq-p+2)+pq}\,,\label{nspower}\\
R &=& \frac{16qp}{\zeta_q}\frac{1-n_s}{p(6q-\sigma q-4)+4}\,.
\label{Rpower}
\ea
As discussed in Sec. \ref{setup}, the tensor-to-scalar ratio $R$ does 
not involve the parameter $\sigma$, since this quantity is invariant 
by taking the noncommutative effect into account; 
this is evident when expressing Eq. (\ref{Rpower}) in terms of $N$.
The main change due to spacetime noncommutativity
appears for the spectral index $n_s$.

For the commutative spacetime ($\sigma=0$) one can easily verify that 
the above results reduce to what was derived in \cite{LS} for the GR 
($q=1, \zeta_1=1$) and RS ($q=2, \zeta_2=2/3$) scenarios. In these 
cases scalar perturbations are red tilted ($n_s<1$). 
The spectrum can be blue tilted
when noncommutativity is switched on.
For example, let us consider the noncommutative limit $\sigma \to 6$. 
In this case we have
\ba \label{nSlimit}
n_s-1=\frac{4(p-1)}{2N(pq-p+2)+pq}\qquad
\text{for}~~\sigma \to 6\,,
\ea
which means $n_s>1$ for $p > 1$. 
Therefore it is possible to explain the loss of 
power in the spectrum at large scales,
as we shall see in the next section.

The exponential potential (\ref{expo}) corresponds to the 
limit $p \to \infty$ in Eqs.~(\ref{nspower}) and (\ref{Rpower}),
thereby yielding 
\ba
n_s-1 &=& \frac{4-(6-\sigma)q}{2N(q-1)+q}\,,\\
R &=& \frac{16q}{\zeta_q[2N(q-1)+q]}\,,
\ea
which is valid for $q\neq 1$.\footnote{The power-law
inflation does not end for the GR case unless the slope of 
the exponential potential changes.}
This gives the border between large-field and hybrid models 
\be \label{border1s}
R = -\frac{16q}{\zeta_q(6q-\sigma q-4)}(n_s-1)\,.
\ee
In the case of GR ($q=1$) we find that the border of large-field and 
hybrid models extends to the region of $n_s>1$ for $\sigma>2$. 
Thus, in the regime where the noncommutative effect becomes 
important ($2 \leq \sigma \leq 6$), one can obtain 
a blue-tilted spectrum even in the large-field models,
which is not possible in the commutative case. 

Note that the border of large-field and small-field models 
corresponds to $p=1$, giving
\be
R = -\frac{16}{\zeta_q(6-\sigma)}(n_s-1)\,.
\ee
This border does not extend to the region $n_s>1$
for $\sigma<6$.


\subsection{The tachyon field $T$}

For the scalar potential (\ref{power}) with the tachyon  
field $T$, we have 
\ba
n_s-1 &=& 
-\frac{p}{6\beta_q^2V_0^q}\frac{pq(4+\theta-\sigma)+4}
{T^{2+qp}}\,,\\
R &=& \frac{16qp^2}{6\beta_q^2\zeta_qV_0^q}T^{-2-qp}\,.
\ea
Since inflation ends at
$T_f^{qp+2}\approx qp^2/(6\beta_q^2V_0^q)$, 
the number of $e$-foldings is estimated as 
\be
N =\frac{3\beta_q^2V_0^q}{p(pq+2)}\,
T^{pq+2}-\frac{qp}{2(pq+2)}\,,\qquad 
p \neq -\frac{2}{q}\,.
\ee
Then we get
\ba
\label{nsT}
n_s-1 &=& -\frac{p(6q-\sigma q-2)+4}{2N(pq+2)+pq}\,,\\
R &=& \frac{16qp}{\zeta_q}\frac{1-n_s}
{p(6q-\sigma q-2)+4}\,,
\label{RT}
\ea
where we used Eq. (\ref{theta}).
For the GR0 spacetime 
($q=1$ and $\sigma=0$), these results reproduce 
what was obtained in Ref.~\cite{SV}.
The tensor-to-scalar ratio is smaller relative to the case of 
the ordinary scalar field $\phi$, thus preferred 
observationally \cite{GST}.
The effect of noncommutativity can lead to a blue-tilted
spectrum ($n_s>1$) as is similar to the case of the 
field $\phi$.

For the exponential potential (\ref{expo}) one gets
\ba
n_s-1 &=& -\frac{4+\theta-\sigma}{2N+1}\,,\\
R &=& \frac{16}{\zeta_q(2N+1)}\,,
\ea
which is obtained by taking the limit $p \to \infty$
in Eqs.~(\ref{nsT}) and (\ref{RT}).
This gives the border of large-/hybrid-field models
\be
R=-\frac{16}{\zeta_q(4+\theta-\sigma)} (n_s-1)\,.
\ee
In the GR case this border belongs to 
the region $n_s>1$ for $\sigma>4$. 

The border of large-/small-field models is 
\be \label{border1t}
R = -\frac{16q}{\zeta_q(6q-\sigma q+2)}(n_s-1)\,,
\ee
which does not extend to the region $n_s>1$
for $\sigma<6$.


\subsection{The difference between $\phi$ and $T$} 

By Eqs.~(\ref{nspower}) and (\ref{nsT}) we find that
the difference between the ordinary field and the tachyon 
field  for the spectral index $n_s$ appears 
both in the denominator and the numerator.
Meanwhile by Eqs.~(\ref{Rpower}) and (\ref{RT})
the difference for the ratio $R$ only appears in 
the denominator
\be 
\label{ratiouni}
R = \frac{16qp}{\zeta_q[2N(qp+2-b)+qp]}\,,
\ee
where $b=p$ for $\vp=\phi$ and $b=0$ for 
$\vp=T$. 
Therefore the tensor-to-scalar ratio in the 
tachyon case is smaller than in the ordinary scalar 
field case when $p>0$. This property implies that the tachyon inflation 
is less affected by an observational pressure as was pointed 
out in the GR commutative case \cite{GST}. 
In the next section we shall study this issue in detail
in the context of noncommutative inflation.


\section{Observational constraints on large-field noncommutative 
models} \label{like2}

In this section we place constraints on large-field noncommutative 
inflationary models using the observational contour bounds obtained 
in Sec.\,\ref{like1}. We plot the theoretical values of $n_s$ and $R$
for $N=45,50,55,60$ on the likelihood contours. 
Typically one can restrict the number of 
$e$-folds to $N \lesssim 65$ \cite{LL2}, but it is sufficient to 
show the values up to $N=60$ to judge whether the models 
we consider are ruled out or not.

Before considering each noncommutative case, it is important to 
understand the basic structure of the theoretical curves on the
$n_s$-$R$ plane.
The effect of the noncommutative parameter $\sigma$ has a 
straightforward geometrical interpretation. 
Let us define $x\equiv n_s-1$ 
and $y \equiv R$, together with the polar coordinates 
$\varrho\equiv\sqrt{x^2+y^2}$ and $\sin\vartheta \equiv y/\varrho$ centered at 
$(1,0)$ in the $n_s$-$R$ plane. 
From the last section, we know that
\ba
y &=& \gamma(q,p,\sigma) x\,,\\
\gamma(q,p,\sigma) &=& -\frac{16qp}
{\zeta_q[p(6q-\sigma q-c)+4]}\,,
\ea
where $c=4$ for $\vp=\phi$ and $c=2$ for $\vp=T$. Then, 
$\varrho^2=(1+\gamma^2)x^2$ and $\tan \vartheta = \gamma$. 
Since $pq>0$ in the cases we consider, $\vartheta$ is a decreasing 
function in terms of $\sigma$.
Therefore, as $\sigma$ increases, 
the theoretical points are rotated clockwise in 
the $\varrho$-$\vartheta$ plane.
This rotation is mainly governed by $\sigma$ rather than 
$p$ when $p$ is large, which can be seen from 
the computation of the logarithmic variation of $\gamma$:
\be \label{rotat}
\frac{d \ln \gamma}{d p} =\frac{4}{qp^2}\frac{d \ln \gamma}{d \sigma}\,.
\ee
This also implies that, for a given $\sigma$, the three models $p=2,4,+\infty$ lie on a wider 
range of radii for smaller values of $q$. As we shall see later, 
this effect is particularly evident in the Gauss-Bonnet case with 
respect to the GR and RS cases in the same (non)commutative class.

The divergence of $\gamma$ at the asymptote $\vartheta=\pi/2$ identifies those models 
generating a scale-invariant scalar spectrum $n_s=1$. They are listed 
in Table \ref{table3} for fixed $\sigma$ and $q$. 
In particular, ordinary scalar class 2 models cannot give $n_s=1$ if one imposes the 
condition $p(q-1)+2\neq 0$ for inflation to have a natural end. The 
tachyonic counterparts are those with $pq+2 \neq 0$, 
and only the GR2 case is excluded.

Note that class 1 patch models admit only one scale-invariant potential
for each inflaton, that is, the linear potential for $\phi$ and the quadratic one for $T$. 
Another frequent case is $p=-2$ (scalar GR0, tachyon GR2 and tachyon GB2), 
which however does not match with the exact power-law solutions of \cite{cal3} 
(RS scalar and GR tachyon). 
Anyway our interest in this paper are the models with positive $p$ ($p \ge 2$)
that lead to natural reheating.

\begin{table}[ht]
\caption{\label{table3}Values of $p$ for scale-invariant models.}
\begin{ruledtabular}
\begin{tabular}{cc|rrr}
       &    &  \multicolumn{3}{c}{$\sigma$} \\
       &    &   0  (Class 0)      &  6  (Class 1)   &   2 (Class 2)   \\ 
\hline
       & GR &   $-2$     &  1 &  $\infty$   \\
$\phi$ & RS &   $-1/2$   &  1    &   $-1$   \\
       & GB &   $\infty$ &  1      &   3   \\ \hline
       & GR &   $-1$     &  2     &   $-2$   \\
$T$    & RS &   $-2/5$   &  2   &   $-2/3$   \\
       & GB &   $-2$     &  2     &   $-6$   \\
\end{tabular}\end{ruledtabular}
\end{table}


\subsection{The ordinary scalar field $\phi$}

Let us first study the observational constraints on the large-field
models for the ordinary field $\phi$.
In Figs.~\ref{fig5}--\ref{fig7} the theoretical values 
(\ref{nspower}) and (\ref{Rpower}) for the potential 
(\ref{power}) are plotted in the GR, RS, and GB cases 
together with $1\sigma$ and $2\sigma$ contour bounds.
Hereafter we shall consider each case separately in order to clarify 
the situation.

\subsubsection{{\rm GR} case}

It is well known that the commutative GR case ($\sigma=0$)
is observationally disfavoured for the 
quartic potential ($p=4$). In this case the theoretical points 
are outside of the $2\sigma$ contour bound for a number of $e$-folds
$N<60$.

\begin{figure}
\includegraphics[height=8.5in,width=3.2in]{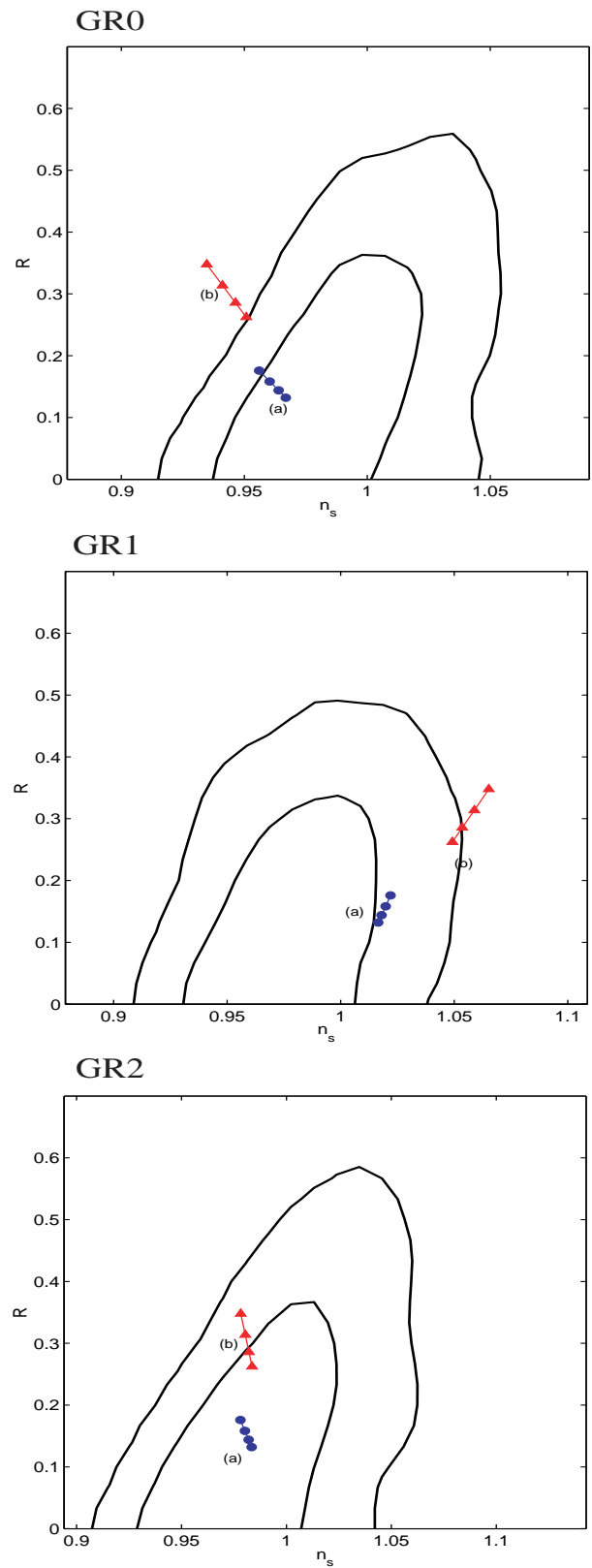}
\caption{\label{fig5}
Observational constraints on large-field models
for the GR ordinary field $\phi$
together with the $1\sigma$ and $2\sigma$ 
contour bounds for three classes
of commutative/noncommutative scenarios.
The theoretical values correspond to 
(\textit{a}) $p=2$ (dots) and (\textit{b}) $p=4$ (triangles),
respectively, with the number of $e$-folds
$N=45, 50, 55, 60$ (from top to bottom).
}
\end{figure}

In the noncommutative class 1 case ($\sigma=6$)
the spectral index $n_s$ is larger than 1 by Eq.~(\ref{nSlimit}).
The tensor-to-scalar ratio $R$ is independent of  $\sigma$, 
so this value is the same as the one in the class 0.
As one can see in Fig.~\ref{fig5}
the quartic potential is outside of
the $2\sigma$ bound for $N<55$. Therefore this case is also 
marginal as in the class 0 case.

The noncommutative class 2 case ($\sigma=2$) corresponds to 
 a scalar spectral index smaller than 1, but it is closer to a
scale-invariant spectrum relative to the class 0 case.
This shifts the theoretical points inside of the $2\sigma$ bound 
and allows the quartic potential even for $N=45$.
Then a ``mild'' spacetime noncommutativity 
in which $\sigma$ is close to 2 is favoured for the 
observational compatibility of the quartic potential.
Note that the quadratic potential ($p=2$) is allowed 
in all three models, irrespective of the degree of noncommutativity,
as clearly shown in Fig.~\ref{fig5}. 

\subsubsection{{\rm RS} case}

In commutative RS spacetime, the quartic potential  is 
under a strong observational pressure as is similar to the GR0 case, and 
the steep inflation driven by an exponential potential
($p \to \infty$) is ruled out \cite{LS,TL}.

\begin{figure}
\includegraphics[height=8.5in,width=3.2in]{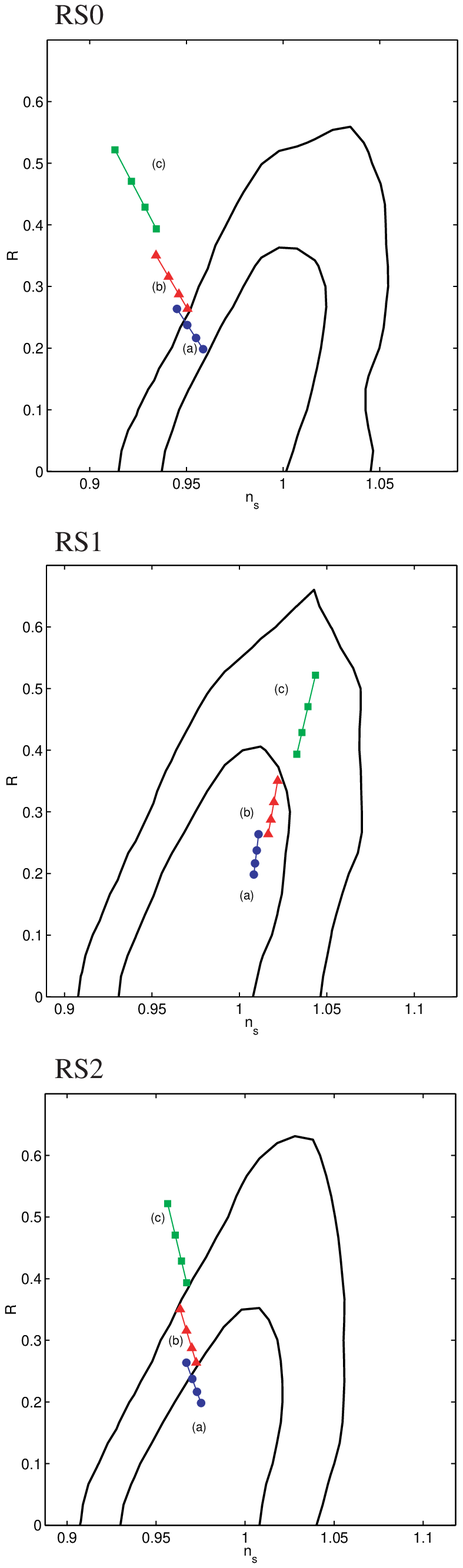}
\caption{\label{fig6}
Observational constraints on large-field models
for the RS ordinary field $\phi$.
Each case corresponds to 
(\textit{a}) $p=2$ (dots), (\textit{b}) $p=4$ (triangles)  
and (\textit{c}) exponential potential with $p \to \infty$ 
(squares), respectively, with the number of $e$-folds
$N=45, 50, 55, 60$ (from top to bottom).
}
\end{figure}

This situation is improved in the class 1 noncommutative scenario. 
Since the spectral index $n_s$ takes a value which is slightly 
larger than 1 and the $2\sigma$ contour bounds extend to 
the region with $R>0.6$, 
even the steep inflation is allowed (see Fig.~\ref{fig6}).

Meanwhile in the class 2 case the exponential 
potential is outside of the $2\sigma$ bound unless the number of 
$e$-folds $N$ is larger than 60.
The quartic potential moves inside of the $2\sigma$ bound relative to 
the class 0 case, thus becoming compatible with observations.

In the RS case strong noncommutativity close to $\sigma=6$ is favoured
observationally rather than mild noncommutativity like $\sigma=2$, 
in contrast with the GR case.

\begin{figure}
\includegraphics[height=8.5in,width=5.0in]{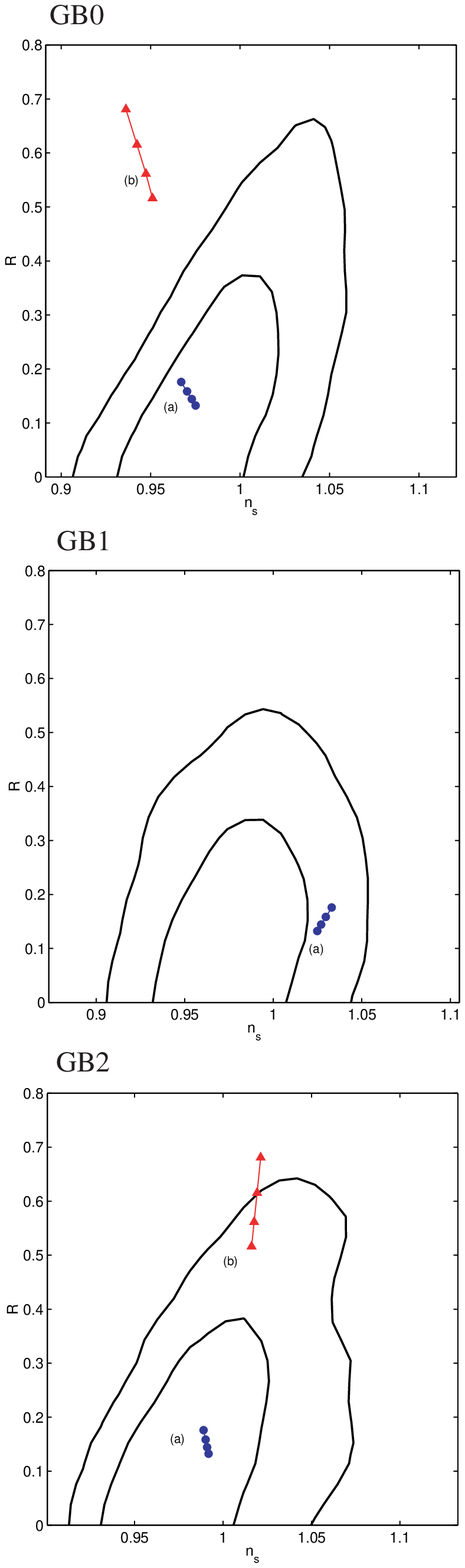}
\caption{\label{fig7}
Observational constraints on large-field models
for the GB ordinary field $\phi$.
Each case corresponds to 
(\textit{a}) $p=2$ (dots) and (\textit{b}) $p=4$ (triangles), 
respectively, with the number of $e$-folds
$N=45, 50, 55, 60$ (from top to bottom).
In the GB1 case the quadratic potential is far 
outside of the $2\sigma$ bound.
}
\end{figure}

\subsubsection{{\rm GB} case}

In the Gauss-Bonnet braneworld cosmology the GB dominant 
stage with $q=2/3$ is followed by the RS stage with $q=2$.
In Ref.~\cite{TSM} theoretical values of $n_s$ and $R$ 
were derived 
for the case where inflation ends in the RS regime.
In this work we study a situation in which the end of 
inflation corresponds to the GB regime.
In this case we do not have a sufficient amount of $e$-folds for $p>6$,
so it is not meaningful to consider steep inflation.

In commutative spacetime the quartic potential  is 
ruled out observationally, while the quadratic potential is 
inside of the $1\sigma$ bound, see Fig.~\ref{fig7}. 

In the class 1 case the spectral index for the quartic model is larger than 1.1
for a number of $e$-folds $N<65$, thus far outside of the $2\sigma$ bound.
In this sense the effect of strong noncommutativity close to $\sigma=6$
is not welcome to save the quartic potential.
On the other hand, the quadratic potential is not ruled out due to a little 
departure from scale invariance.

The class 2 noncommutative scenario exhibits an interesting 
feature to have $n_s$ close to 1 even for the quartic potential.
As seen in Fig.~\ref{fig7} the quartic potential is within 
the $2\sigma$ bound for $N>50$, thereby compatible with 
observations. This situation is similar to the GR case.


\subsection{The tachyon field $T$}

Let us next consider the observational constraint 
on the tachyonic large-field models. 

\subsubsection{{\rm GR} case}

The GR commutative case was already investigated in 
Refs.~\cite{SV,GST}. Since the tensor-to-scalar ratio 
is smaller relative to the normal scalar field case,
this leads to the compatibility with observations.
Even steep inflation is 
deep within the $2\sigma$ contour bound.

\begin{figure}
\includegraphics[height=8.5in,width=3.2in]{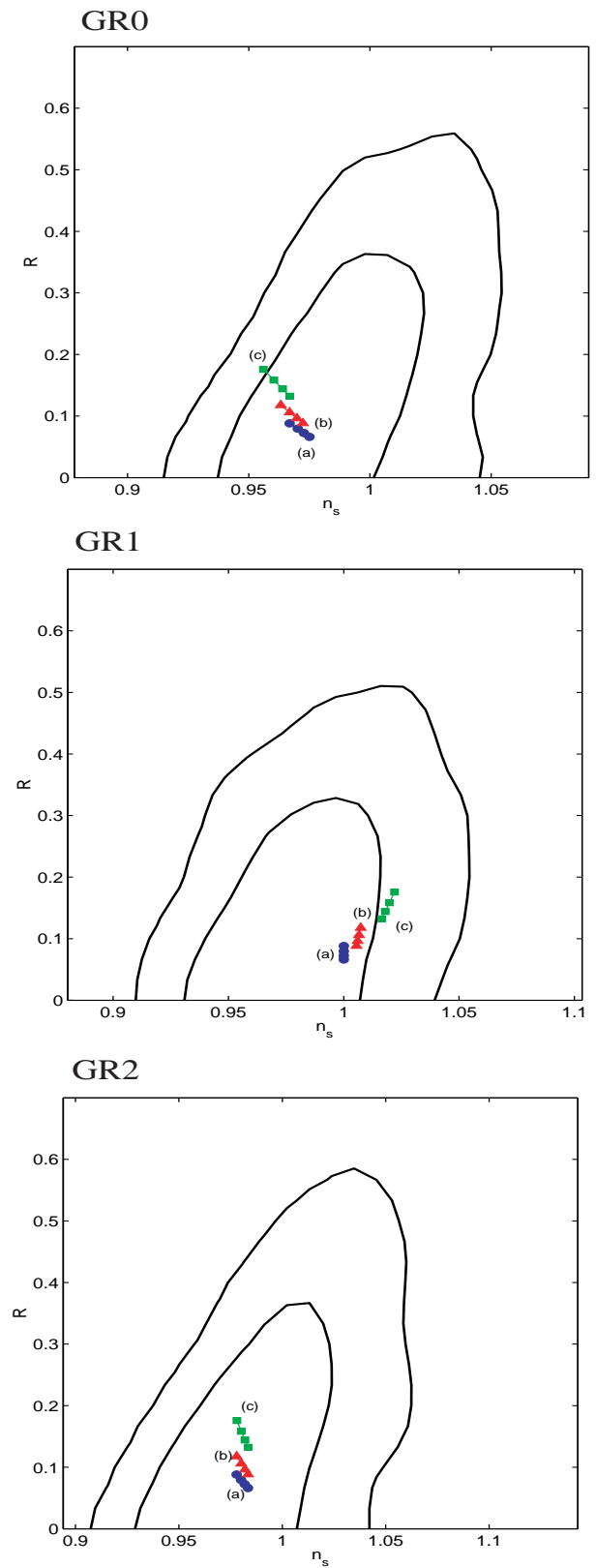}
\caption{\label{fig8}
Observational constraints on large-field models
for the GR tachyon field $T$
together with the $1\sigma$ and $2\sigma$ 
contour bounds for three classes
of commutative/noncommutative scenarios.
Each case corresponds to 
(\textit{a}) $p=2$ (dots), (\textit{b}) $p=4$ (triangles)  
and (\textit{c}) exponential potential with $p \to \infty$ 
(squares), respectively, with the number of $e$-folds
$N=45, 50, 55, 60$ (from top to bottom).
}
\end{figure}

Because of this small value of $R$, the class 1 and class 2
noncommutative scenarios are also allowed as shown
in Fig.~\ref{fig8}.  
The class 1 scenario corresponds to a spectral index $n_s$
larger than 1, but this does not deviate from a
scale-invariant spectrum.
All cases with $p=2$, $p=4$ and $p=\infty$ are inside of 
the $2\sigma$ contour bound.

\subsubsection{{\rm RS} case}

\begin{figure}
\includegraphics[height=8.5in,width=3.2in]{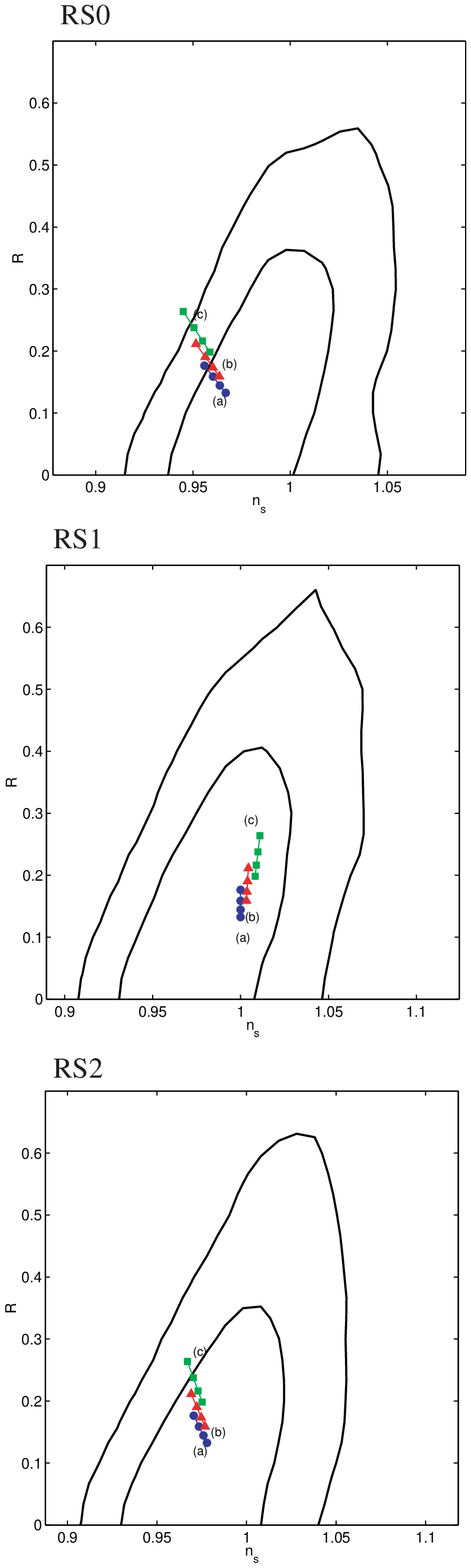}
\caption{\label{fig9}
Observational constraints on large-field models
for the RS tachyon field $T$.
Each case corresponds to 
(\textit{a}) $p=2$ (dots), (\textit{b}) $p=4$ (triangles)  
and (\textit{c}) exponential potential with $p \to \infty$ 
(squares), respectively, with the number of $e$-folds
$N=45, 50, 55, 60$ (from top to bottom).
}
\end{figure}

The RS case exhibits larger values of the tensor-to-scalar
ratio compared to the GR case. However, the quadratic and 
quartic potentials are always within the $2\sigma$
bound. The exponential potential is also allowed
for the $e$-folds with $N \gtrsim 50$. See Fig. \ref{fig9}.

\subsubsection{{\rm GB} case}

\begin{figure}
\includegraphics[height=8.5in,width=3.2in]{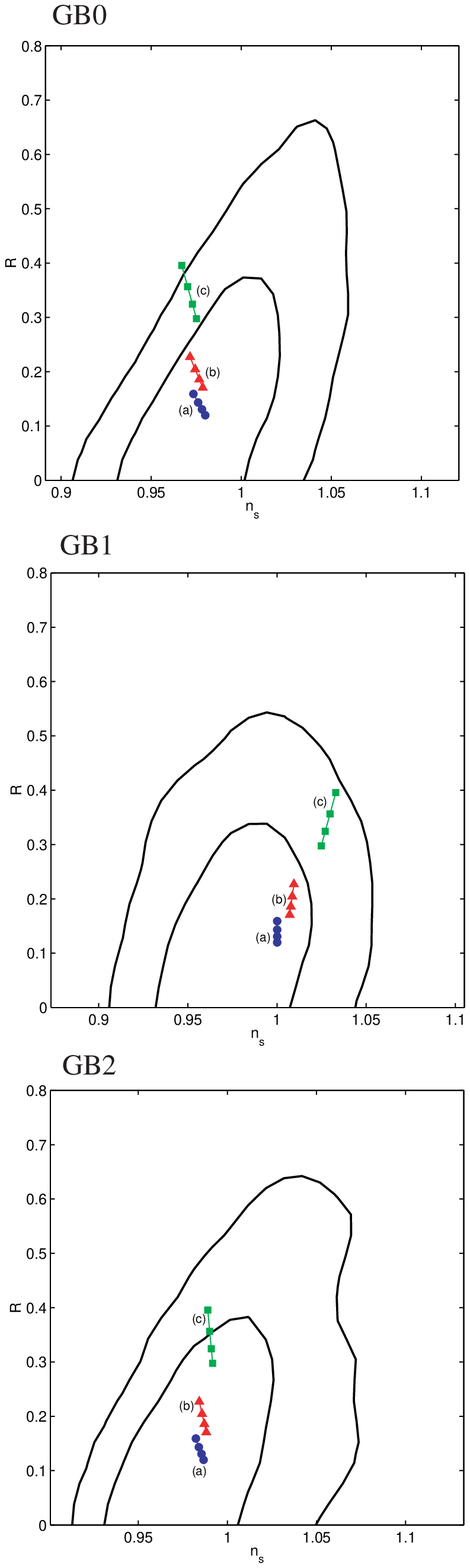}
\caption{\label{fig10}
Observational constraints on large-field models
for the GB tachyon field $T$.
Each case corresponds to 
(\textit{a}) $p=2$ (dots), (\textit{b}) $p=4$ (triangles)  
and (\textit{c}) exponential potential with $p \to \infty$ 
(squares), respectively, with the number of $e$-folds
$N=45, 50, 55, 60$ (from top to bottom).
}
\end{figure}

By Eq.~(\ref{rotat}) each inflationary model ($p=2, 4, \infty$)
in the GB case ($q=2/3$) lies on a wider range of radii $\varrho$
relative to the GR and RS cases.
In spite of this property, even steep inflation is compatible 
with observations in both commutative and noncommutative 
spacetimes.
In summary, tachyon inflation is allowed irrespective of 
the slope of the potential due to a small tensor-to-scalar ratio in all patch cosmologies we have considered. See Fig. \ref{fig10}.

\subsection{Suppression of CMB low multipoles in noncommutative inflation}

In Refs.~\cite{HL1,TMB,HL2} it was shown that it is possible 
to explain the loss of the power spectrum at low multipoles at least partially 
using the modified spectrum in the UV regime ($\tau \gg kl_s^2$).
Here we will consider the situation in which
the spectrum on cosmologically relevant scales 
is generated in the IR noncommutative regime.

As seen in the likelihood contours, the best-fit value of 
$n_s$ is smaller than 1 and is 
insensitive to which prescription we adopt for the spacetime 
(non)commutative structure.
The loss of power on largest scales is difficult to 
be explained in the standard concordance scenario.
If we take the effect of spacetime noncommutativity
into account, it is possible to have a suppression of power
due to a blue-tilted spectrum.
For example, the potential (\ref{power}) gives rise
to the blue spectrum for the GR1 noncommutative case.
Of course, the large spectral index $n_s \gtrsim 1.05$
is ruled out as seen in Fig.~\ref{fig5}, but 
the quadratic potential ($p=2$) gives the observationally 
allowed value around $n_s \sim 1.02$.
The quartic potential ($p=4$) corresponds to a marginal 
compatibility with observations,
but it is welcome to explain the loss of power 
on the largest scales.

\begin{figure}
\includegraphics[height=3.5in,width=3.5in] {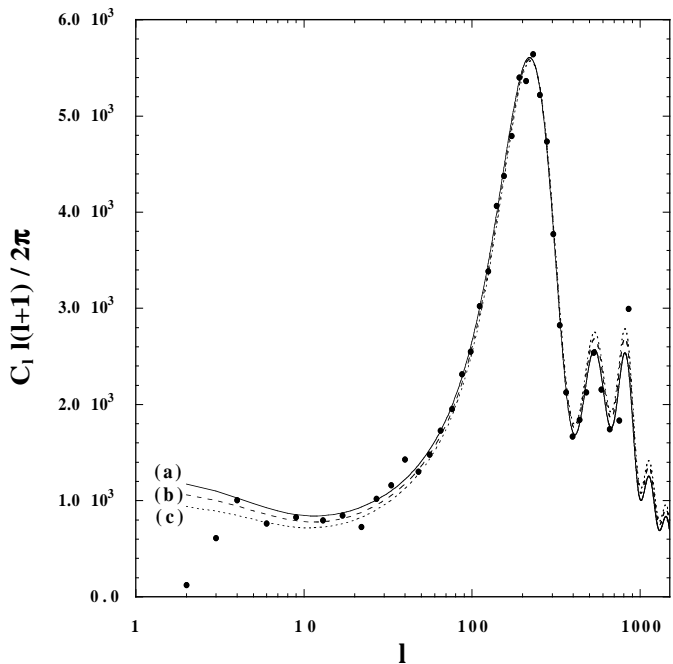}
\caption{\label{fig11}
The CMB angular power spectrum showing the effects of
suppression of power at low multipoles. 
Curve~(\textit{a}) is the GR commutative model
with $(n_s, R)=(0.967, 0.132)$
corresponding to the quadratic potential.
Curves~(\textit{b}) and (\textit{c}) are 
the GR1 noncommutative scenario with
$(n_s, R)=(1.018, 0.144)$ and 
$(n_s, R)=(1.049, 0.263)$, respectively.
Note that these values
are achieved for the quadratic 
and quartic potentials in the GR1 case, respectively. 
}
\end{figure}

In Fig.~\ref{fig11} we plot the CMB angular 
power spectra for several different cases.
The spectrum exhibits some suppression around 
$1<l\lesssim 10$ in noncommutative spacetime 
relative to the commutative one.
The quartic potential leads to a stronger suppression
compared to the quadratic one, but 
the smaller-scale spectrum tends to show some disagreement 
with observations for larger $n_s$.
Anyway, it is intriguing that single-field noncommutative 
inflation leads to a blue-tilted spectrum 
suitable for explaining the 
loss of power at low multipoles, since this is difficult to 
be achieved in commutative spacetime 
unless we introduce another scalar field as in the case of 
hybrid inflation (see also \cite{LMMR}). 

\section{Conclusion and discussion} \label{disc}

In this paper we have investigated observational constraints on  
a number of patch noncommutative inflationary scenarios including 
general relativity, Randall-Sundrum and 
Gauss-Bonnet braneworld.
We expressed the inflationary observables \{$A_s^2$, $R$, $n_s$, $n_t$,
$\alpha_s$, $\alpha_t$\} in terms of horizon-flow parameters
both for a normal scalar field $\phi$ and a tachyon field $T$.
We showed that the likelihood analysis of the observables is the same 
for both types of scalar fields by using HF parameters.
It is known that the consistency relation between 
the tensor spectral index and the tensor-to-scalar ratio is the same 
both in commutative GR and RS cases.
This property also holds in induced-gravity braneworld 
inflation \cite{BMW} and 
in generalized Einstein theories including 
four-dimensional dilaton gravity and scalar-tensor theories \cite{TG04}.
The spacetime noncommutativity breaks this degeneracy and 
provides a variety of consistency relations [see Eq.~(\ref{ntconeq}) and Table \ref{table2}].
We have considered two classes of noncommutative models
and evaluated the power spectra in the IR limit.
The strength of noncommutativity is measured by a parameter
$\sigma$, and we classified the models into three classes: (i) commutative 
spacetime with $\sigma=0$, (ii) the class 1 IR case with $\sigma=6$
and (iii) the class 2 IR case with $\sigma=2$.
The perturbations are always blue-tilted for the class 1 scenario,
thus giving positive values of $n_t/R$.
This unusual property comes from the fact that the mechanism for
generating fluctuations is different from the standard case 
due to the existence of the stringy uncertainty relation.

We carried out likelihood analyses in terms of inflationary and cosmological
parameters using the data set coming from WMAP, the 2dF, and 
SDSS galaxy redshift surveys.
The numerical analysis showed that both the HF parameter $\epsilon_3$ 
and the slow-roll parameter $\xi^2$ are 
poorly constrained and can be consistently set equal to zero. 
We ran the code by varying the parameter $\sigma$
in the range $0 \le \sigma \le 6$ and found that $\sigma$ does not 
show a good convergence. 
This means that current observations do not choose a preferred commutative 
or noncommutative model. 
We then performed a likelihood analysis for three fixed values of 
$\sigma$ ($\sigma=0, 2, 6$).
As seen in Figs.~\ref{fig2}--\ref{fig4} one can find 
some difference in the $n_s$-$R$ plane by the modification of consistency 
relations due to noncommutative spacetime.
The main change appears in the maximum value of $R$ ($=R_\text{max}$)
and it ranges in the region $0.5 \lesssim R_\text{max} \lesssim 0.7$.

We also placed constraints on the large-field monomial potential (\ref{power})
for noncommutative GR/RS/GB scenarios. We found the following 
interesting results. 

\vspace{0.2cm}

For the ordinary scalar field $\phi$:

\begin{itemize}
\item The quartic potential is rescued 
from the marginal rejection in the noncommutative class 2 GR case
($\sigma=2$).

\item Steep inflation driven by an exponential potential 
is allowed in the noncommutative class 1 RS case ($\sigma=6$).
The quartic potential is compatible with observations both in the class 1
and class 2 RS cases, but it is not so in the RS commutative case.

\item The quartic potential 
exhibits a compatibility with observations for the class 2 GB case,
while it does not in the other two cases (GB0 and GB1).
\end{itemize}

For the tachyon field $T$:

\begin{itemize}
\item A scale-invariant spectrum ($n_s=1$) 
is generated for $p=2$ in the noncommutative class 1 case 
irrespective of the kind of patch cosmologies.

\item Even steep inflation is allowed 
due to small values of the tensor-to-scalar ratio in three classes
of patch cosmologies.
\end{itemize}
All these properties have been investigated both analytically and numerically.
We also pointed out the possibility to explain the suppression of CMB 
low multipoles using a blue-tilted spectrum generated in the IR regime of
noncommutative spacetime. Note that this is different from the approach in 
Ref.~\cite{TMB} in which an  
intermediate spectrum between the IR and UV regions
is used to explain the loss of power. 
Although noncommutativity can provide a better fit of the spectrum for low multipoles,
it is not easy to explain the loss of power at $\ell=2,3$. 
The suppression in this region, corresponding to the Sachs-Wolfe plateau, 
is difficult to be achieved even with a very blue-tilted spectrum 
$n_s \gtrsim 3$ \cite{PTZ} especially for $\ell=2$.
This is still an open issue and would deserve further study.

It would be interesting to investigate whether some regions 
of the line of patches $\theta$ are excluded or not by observations. 
A clear answer in this respect would constrain any new braneworld 
scenario with a nonstandard Friedmann equation with $\theta\neq 0, \pm 1$. 
In the GR case ($\theta=0$), we have addressed a similar question 
for $\sigma$ and performed a likelihood analysis with a very large prior 
($|\sigma|<100$). The parameter did not show a good convergence, 
since the tensor index $n_t$ can be made smaller by choosing 
a smaller $R$ in Eq. (\ref{ntconeq}). 
Since the same result holds when varying the set 
$\{\theta,n_t, \zeta_qR\}$, one has to consider \emph{fixed} 
values of any extra parameter which modifies the four-dimensional scenario.
Then we cannot say anything \emph{a priori} about the viability of a general 
patch cosmology.

A fundamental question to be answered is: What about cosmic 
confusion? Can we rely on the consistency equations as a smoking gun 
for both braneworld and noncommutative scenarios? 
The answer is presumably no, since, as it typically happens in cosmology, other 
completely different frameworks could mimic the features we have 
exploited. With this sky tolerance, even simple 4D multifield 
configurations produce a nonstandard set of consistency relations 
(See Refs.~\cite{BMR,WBMR,TPB}). 
The subject has to be further explored in a 
more precise way than that provided by the patch formalism
in order to find out more characteristic and sophisticated predictions.


\section*{ACKNOWLEDGMENTS}
It is a pleasure to thank Robert Brandenberger, Andrew Liddle, 
Roy Maartens, and M. Sami for useful discussions.
We also thank Sam Leach for providing the SDSS code.
S.T. thanks the universities of Sussex, Queen Mary, and Portsmouth  
for supporting their visits and warm hospitality 
during various stages of completion of this work.


\begin{thebibliography}{75}

\bibitem{ben03}  
C.L. Bennett \textit{et al.}, Astrophys. J., Suppl. 
Ser. \textbf{148}, 1 (2003) [arXiv:astro-ph/0302207].

\bibitem{spe03}  
D.N. Spergel \textit{et al.}, Astrophys. J., Suppl. 
Ser. \textbf{148}, 175 (2003) [arXiv:astro-ph/0302209].

\bibitem{pei03}  
H.V. Peiris \textit{et al.}, Astrophys. J., Suppl. Ser.
\textbf{148}, 213 (2003) [arXiv:astro-ph/0302225].

\bibitem{bri03}  
S.L. Bridle, A.M. Lewis, J. Weller, and G. 
Efstathiou, Mon. Not. R. Astron. Soc. \textbf{342}, L72 (2003) 
[arXiv:astro-ph/0302306].

\bibitem{BLM}
V. Barger, H.S. Lee, and D. Marfatia, 
Phys. Lett. B \textbf{565}, 33 (2003) [arXiv:hep-ph/0302150].

\bibitem{KKMR}
W.H. Kinney, E.W. Kolb, A. Melchiorri, and A. Riotto, 
Phys. Rev. D \textbf{69}, 103516 (2004) [arXiv:hep-ph/0305130].

\bibitem{LL}     
S.M. Leach and A.R. Liddle, Phys. Rev. D 
\textbf{68}, 123508 (2003) [arXiv:astro-ph/0306305].

\bibitem{teg04} 
SDSS collaboration, M. Tegmark \textit{et al.}, Phys. 
Rev. D \textbf{69}, 103501 (2004) [arXiv:astro-ph/0310723].

\bibitem{planc}  
http://astro.estec.esa.nl/SA-general/Projects/Planck/

\bibitem{RS}
L. Randall and R. Sundrum, 
Phys.\ Rev.\ Lett.\  \textbf{83}, 4690 (1999) [arXiv:hep-th/9906064].

\bibitem{mar03}  
R. Maartens, Living Rev. Relativity \textbf{7}, 1 [arXiv:gr-qc/0312059].

\bibitem{BVD}    
P. Brax, C. van de Bruck, and A.-C. Davis, 
arXiv:hep-th/0404011.

\bibitem{cal3}   
G. Calcagni, Phys. Rev. D \textbf{69}, 103508 (2004) 
[arXiv:hep-ph/0402126].

\bibitem{MWBH}
R.~Maartens, D.~Wands, B.A.~Bassett, and I.P.C.~Heard, 
Phys.\ Rev.\ D \textbf{62}, 041301 (2000)
[arXiv:hep-ph/9912464].

\bibitem{LMW}
D. Langlois, R. Maartens, and D. Wands,
Phys. Lett. B \textbf{489}, 259 (2000) [arXiv:hep-th/0006007].

\bibitem{HuL1}   
G. Huey and J.E. Lidsey, Phys. Lett. B \textbf{514}, 
217 (2001) [arXiv:astro-ph/0104006].

\bibitem{HuL2}   
G. Huey and J.E. Lidsey, Phys. Rev. D \textbf{66}, 
043514 (2002) [arXiv:astro-ph/0205236].

\bibitem{LS}     
A.R. Liddle and A.J. Smith, Phys. Rev. D 
\textbf{68}, 061301 (2003) [arXiv:astro-ph/0307017].

\bibitem{TL}     
S. Tsujikawa and A.R. Liddle, J. Cosmol. Astropart. 
Phys. \textbf{03}, 001 (2004) [arXiv:astro-ph/0312162].

\bibitem{LN}     
J.E. Lidsey and N.J. Nunes, Phys. Rev. D 
\textbf{67}, 103510 (2003) [arXiv:astro-ph/0303168].

\bibitem{DLMS}
J.F.~Dufaux, J.E.~Lidsey, R.~Maartens, and M.~Sami, Phys. Rev. D \textbf{70}, 083525 (2004) [arXiv:hep-th/0404161].

\bibitem{TSM}    
S. Tsujikawa, M. Sami, and R. Maartens, Phys. Rev. D \textbf{70}, 063525 (2004) [arXiv:astro-ph/0406078].

\bibitem{ABM}
S.~Alexander, R.~Brandenberger, and J.~Magueijo, 
Phys.\ Rev.\ D \textbf{67}, 081301 (2003)
[arXiv:hep-th/0108190].

\bibitem{BH}     
R. Brandenberger and P.-M. Ho, Phys. Rev. D 
\textbf{66}, 023517 (2002) [arXiv:hep-th/0203119].

\bibitem{HL1}    
Q.-G. Huang and M. Li, J. High Energy Phys. 
\textbf{06}, 014 (2003) [arXiv:hep-th/0304203].

\bibitem{FKM}
M.~Fukuma, Y.~Kono, and A.~Miwa,
Nucl.\ Phys.\ \textbf{B682}, 377 (2004)
[arXiv:hep-th/0307029].

\bibitem{FKM2}   
M. Fukuma, Y. Kono, and A. Miwa, \eprint{arXiv:hep-th/0312298}.

\bibitem{TMB}    
S. Tsujikawa, R. Maartens, and R. Brandenberger, 
Phys. Lett. B \textbf{574}, 141 (2003) [arXiv:astro-ph/0308169].

\bibitem{HL2}    
Q.-G. Huang and M. Li, J. Cosmol. Astropart. Phys. 
\textbf{11}, 001 (2003) [arXiv:astro-ph/0308458].

\bibitem{HL3}    
Q.-G. Huang and M. Li, arXiv:astro-ph/0311378.

\bibitem{LiLi}
D.-J. Liu and X.-Z. Li,
arXiv:astro-ph/0402063.

\bibitem{KLM}
H.S. Kim, G.S. Lee, and Y.S. Myung,
arXiv:hep-th/0402018.

\bibitem{KLLM}
H.S. Kim, G.S. Lee, H.W. Lee, and Y.S. Myung,
arXiv:hep-th/0402198.

\bibitem{Cai}
R.-G. Cai,
Phys.\ Lett.\ B \textbf{593}, 1 (2004)
[arXiv:hep-th/0403134].

\bibitem{LMMP}   
F. Lizzi, G. Mangano, G. Miele, and M. Peloso, J. 
High Energy Phys. \textbf{06}, 049 (2002) [arXiv:hep-th/0203099].

\bibitem{AN}
S.A. Alavi and F. Nasseri,
arXiv:astro-ph/0406477.

\bibitem{cal4}   
G. Calcagni, Phys. Rev. D to appear [arXiv:hep-th/0406006].

\bibitem{cal5}   
G. Calcagni, arXiv:hep-ph/0406057.

\bibitem{myu04}  
Y.S. Myung, arXiv:hep-th/0407066.

\bibitem{KM}     
K.H. Kim and Y.S. Myung, arXiv:astro-ph/0406387.

\bibitem{gib02}  
G.W. Gibbons, Phys. Lett. B \textbf{537}, 1 (2002) [arXiv:hep-th/0204008].

\bibitem{CGJP}   
D. Choudhury, D. Ghoshal, D.P. Jatkar, and S. Panda, 
Phys. Lett. B \textbf{544}, 231 (2002) [arXiv:hep-th/0204204].

\bibitem{KL}     
L. Kofman and A.D. Linde, J. High Energy Phys. 
\textbf{07}, 004 (2002) [arXiv:hep-th/0205121].

\bibitem{SCQ}
M. Sami, P. Chingangbam, and T. Qureshi,
Phys.\ Rev.\ D \textbf{66}, 043530 (2002) [arXiv:hep-th/0205179].

\bibitem{FKS}    
A. Frolov, L. Kofman, and A. Starobinsky, Phys. 
Lett. B \textbf{545}, 8 (2002) [arXiv:hep-th/0204187].

\bibitem{BBS}    
M.C. Bento, O. Bertolami, and A.A. Sen, Phys. Rev. D 
\textbf{67}, 063511 (2003) [arXiv:hep-th/0208124].

\bibitem{BSS}    
M.C. Bento, N.M.C. Santos, and A.A. Sen, 
arXiv:astro-ph/0307292.

\bibitem{PS}     
B.C. Paul and M. Sami, Physical Review D to appear, 
arXiv:hep-th/0312081.

\bibitem{rae04}  
J. Raeymaekers, arXiv:hep-th/0406195.

\bibitem{PCZZ}   Y.-S. Piao, R.-G. Cai, X. Zhang, and Y.-Z. Zhang, Phys. Rev. D \textbf{66}, 121301(R) (2002) [arXiv:hep-ph/0207143].

\bibitem{GST}    
M.R. Garousi, M. Sami, and S. Tsujikawa, Phys. Rev. D \textbf{70}, 043536 (2004) [arXiv:hep-th/0402075].

\bibitem{KKLT}
S.~Kachru, R.~Kallosh, A.~Linde, and S.P.~Trivedi,
Phys.\ Rev.\ D \textbf{68}, 046005 (2003)
[arXiv:hep-th/0301240].

\bibitem{STG}    
D.J. Schwarz, C.A. Terrero-Escalante, and A.A. 
Garc\'{\i}a, Phys. Lett. B \textbf{517}, 243 (2001) 
[arXiv:astro-ph/0106020].

\bibitem{kin02}  
W.H. Kinney, Phys. Rev. D \textbf{66}, 083508 (2002) 
[arXiv:astro-ph/0206032].

\bibitem{LCS}    
A. Lewis, A. Challinor, and A. Lasenby, Astrophys. 
J. \textbf{538}, 473 (2000) [arXiv:astro-ph/9911177].

\bibitem{LB}     
A. Lewis and S. Bridle, Phys. Rev. D \textbf{66}, 
103511 (2002) [arXiv:astro-ph/0205436].

\bibitem{camb}   
http://camb.info/

\bibitem{wmap}   
http://lambda.gsfc.nasa.gov/

\bibitem{per01}  
W.J. Percival \textit{et al.}, Mon. Not. R. Astron. 
Soc. \textbf{327}, 1297 (2001) [arXiv:astro-ph/0105252].

\bibitem{CBI}
T.J. Pearson {\it et al.}, Astrophys. J. \textbf{591}, 556 (2003) [arXiv:astro-ph/0205388].

\bibitem{VSA}
K. Grainge {\it et al.}, Mon. Not. R. Astron. Soc. \textbf{341}, L23 (2003) [arXiv:astro-ph/0212495].

\bibitem{ACBAR}
C.L. Kuo {\it et al.}, Astrophys. J. \textbf{600}, 32 (2004) [arXiv:astro-ph/0212289].

\bibitem{SV}   
D.A. Steer and F. Vernizzi, Phys. Rev. D \textbf{70}, 043527 (2004) [arXiv:hep-th/0310139].

\bibitem{LL2}    
A.R. Liddle and S.M. Leach, Phys. Rev. D 
\textbf{68}, 103503 (2003) [arXiv:astro-ph/0305263].

\bibitem{LMMR}   
M. Liguori, S. Matarrese, M. Musso, and A. Riotto, arXiv:astro-ph/0405544.

\bibitem{BMW}
M. Bouhmadi-L\'{o}pez, R. Maartens, and D. Wands,
arXiv:hep-th/0407162.

\bibitem{TG04}   
S. Tsujikawa and B. Gumjudpai,
Phys. Rev. D \textbf{69}, 123523 (2004)
[arXiv:astro-ph/0402185].

\bibitem{PTZ}
Y.-S. Piao, S. Tsujikawa, and X. Zhang, Class. Quantum Grav. \textbf{21}, 4455 (2004) [arXiv:hep-th/0312139]. 

\bibitem{BMR}
N. Bartolo, S. Matarrese, and A. Riotto,
Phys. Rev. D \textbf{64}, 123504 (2001) [arXiv:astro-ph/0107502].

\bibitem{WBMR}   
D. Wands, N. Bartolo, S. Matarrese, and A. Riotto, 
Phys. Rev. D \textbf{66}, 043520 (2002) [arXiv:astro-ph/0205253].

\bibitem{TPB}
S. Tsujikawa, D. Parkinson, and B.A. Bassett,
Phys. Rev. D \textbf{67}, 083516 (2003) [arXiv:astro-ph/0210322].

\end{thebibliography}
\end{document}